\documentclass[]{mosi}

\usepackage{helvet}

\usepackage{amsmath} 
\usepackage{natbib}
\usepackage{graphicx}
\usepackage{subcaption} 

\usepackage[toc,page,header]{appendix}
\usepackage[utf8]{inputenc} 
\usepackage[T1]{fontenc}    
\usepackage{hyperref}       
\usepackage{url}            
\usepackage{booktabs}       

\usepackage{amsfonts}       
\usepackage{nicefrac}       
\usepackage{microtype}      
\usepackage{wrapfig}

\usepackage{amssymb}        

\usepackage{titletoc}
\usepackage{minitoc}

\usepackage{array}
\usepackage{etoolbox}

\definecolor{lightblue}{RGB}{200, 230, 255}  
\definecolor{headerblue}{RGB}{150, 200, 255} 

\usepackage{pgfplots}
\usepackage{xcolor}
\usepackage{float}
\usepackage{comment}
\usepackage{multirow}
\usepackage{makecell}
\usepackage{siunitx}
\usepackage{tikz}
\usepackage{pgf-pie}
\usepackage[export]{adjustbox}

\usepackage{ragged2e}
\usepackage{tabularx}
\usepackage{caption}
\usepackage{enumitem}
\usepackage{pifont}
\usepackage[hang,flushmargin]{footmisc}

\usepackage{tcolorbox}
\tcbuselibrary{breakable}
\tcbuselibrary{skins}

\usepackage{tabularx}
\usepackage{listings}

\usepackage{threeparttable}
\usepackage{setspace}

\usepackage[scheme=plain]{ctex}

\usepackage[table]{xcolor}
\usepackage{tabularx}
\usepackage{array}
\usepackage{multirow}
\usepackage{booktabs}
\usepackage{wrapfig} 

\newcommand{\up}{$\uparrow$}
\newcommand{\down}{$\downarrow$}
\newcommand{\bnum}[1]{\textbf{#1}}

\definecolor{oursgray}{gray}{0.95}

\definecolor{MossCyan}{HTML}{82D9FF} 
\definecolor{MossBlue}{HTML}{82B1FF}

\definecolor{ForestGreen}{RGB}{34, 139, 34}
\definecolor{Red}{RGB}{255, 0, 0}

\definecolor{tickG}{HTML}{00C853}  

\definecolor{crossR}{HTML}{FF1744} 

\newcommand{\cmark}{\textcolor{tickG}{\bfseries\ding{52}}}
\newcommand{\xmark}{\textcolor{crossR}{\bfseries\ding{56}}}

\definecolor{frenchblue}{rgb}{0.0, 0.45, 0.73}
\definecolor{babyblue}{rgb}{0.54, 0.81, 0.94}
\definecolor{classicrose}{rgb}{0.98, 0.8, 0.91}
\definecolor{beige}{rgb}{0.96, 0.96, 0.86}
\definecolor{forestgreen}{HTML}{2e7d43}

\definecolor{blue1}{HTML}{91BBE6}
\definecolor{blue2}{HTML}{3F90E0}
\definecolor{blue3}{HTML}{316FAD}

\definecolor{color1}{HTML}{FF9999}
\definecolor{color2}{HTML}{FF6666}
\definecolor{color3}{HTML}{FF3333}
\definecolor{color4}{HTML}{E60000}
\definecolor{color5}{HTML}{B30000}
\definecolor{color6}{HTML}{8CD98C}
\definecolor{color7}{HTML}{53c653}
\definecolor{color8}{HTML}{00B050}
\definecolor{color9}{HTML}{2d862d}
\definecolor{color10}{HTML}{206020}
\definecolor{color11}{HTML}{cca300}

\newtcolorbox{promptbox}[2][]{
    colback=white,
    coltext=black,
    arc=3mm,
    boxrule=0.5pt,
    colframe=black!60!white,
    title={#2},
    colbacktitle=black,
    coltitle=white,
    fonttitle=\bfseries,
    top=8pt,
    bottom=8pt,
    left=10pt,
    right=10pt,
    breakable,
    before upper={
        \linespread{1}\selectfont
        \setlength{\parskip}{1ex plus 0.2ex minus 0.2ex}
        \setlength{\parindent}{0pt}
    },
    #1
}

\title{MOSS-Audio-Tokenizer: Scaling Audio Tokenizers for Future Audio Foundation Models}

\author{MOSI.AI\textsuperscript{*}}

\abstract{

Discrete audio tokenizers are fundamental to empowering large language models with native audio processing and generation capabilities. Despite recent progress, existing approaches often rely on pretrained encoders, semantic distillation, or heterogeneous CNN-based architectures. These designs introduce fixed inductive biases that limit reconstruction fidelity and hinder effective scaling. In this paper, we argue that discrete audio tokenization should be learned fully end-to-end using a homogeneous and scalable architecture.
To this end, we first propose \textbf{CAT} (\textbf{C}ausal \textbf{A}udio \textbf{T}okenizer with \textbf{T}ransformer), a purely Transformer-based architecture that jointly optimizes the encoder, quantizer, and decoder from scratch for high-fidelity reconstruction. Building on the CAT architecture, we develop MOSS-Audio-Tokenizer, a large-scale audio tokenizer featuring 1.6 billion parameters, pre-trained on 3 million hours of diverse, general audio data. We show that this simple, fully end-to-end approach built from homogeneous, causal Transformer blocks scales gracefully and supports high-fidelity reconstruction across diverse audio domains.
Across speech, sound, and music, MOSS-Audio-Tokenizer consistently outperforms prior codecs over a wide range of bitrates, while exhibiting predictable improvements with increased scale. Notably, leveraging the discrete tokens from our model, we develop the first purely autoregressive TTS model that surpasses prior non-autoregressive and cascaded systems. Furthermore, MOSS-Audio-Tokenizer enables competitive ASR performance without auxiliary encoders. Our findings position the CAT architecture as a unified, scalable interface for the next generation of native audio foundation models.

}

\checkdata[Code]{\url{https://github.com/OpenMOSS/MOSS-Audio-Tokenizer}}
\checkdata[Model]{\url{https://huggingface.co/OpenMOSS-Team/MOSS-Audio-Tokenizer}}

\begin{document}
\maketitle
\begingroup
\renewcommand{\thefootnote}{\fnsymbol{footnote}}
\setcounter{footnote}{1}
\footnotetext{Full contributors can be found in the Contributors section.}
\endgroup

% ===== Original meta lines (content unchanged) =====
% \begin{spacing}{1.0}
% {\small \noindent \textbf{Date:} December 31, 2025 \par}
% {\small \noindent \textbf{Demo:} \url{https://openmoss.github.io/MOSS-Transcribe-Diarize-demo/} \par}
% {\small \noindent \textbf{Huggingface Space:} \url{https://huggingface.co/spaces/OpenMOSS-Team/MOSS-transcribe-diarize} \par}
% \end{spacing}

% ===== Original content starts here (sections/tables/figures unchanged) =====

\section{Introduction}

% \begin{figure}[t]
% \vspace{-14pt}
%   \centering
%   \includegraphics[width=0.7\linewidth]{Images/all_tokenizers_sim_vs_bps_v2.pdf}
% \caption{Audio reconstruction quality comparison between Cat and open-source audio tokenizers. 
% At comparable bitrates (0--4000 bps), Cat consistently achieves higher SIM scores than all other compared tokenizers.}
%   \label{fig:all_tokenizers_sim_vs_bps}
% \vspace{-10pt}
% \end{figure}

\begin{table*}[t]
  \centering
  \fontsize{8pt}{10pt}\selectfont
  \setlength{\tabcolsep}{4pt}
  \renewcommand{\arraystretch}{1.15}
\caption{Comparison of representative audio tokenizers with respect to architectural design and functional capabilities.
\cmark\ indicates support, \xmark\ indicates not supported, and `--' indicates not specified.
Trans.\ denotes Transformer, and Hybrid denotes a hybrid architecture combining CNN and Transformer. End-to-end optimize indicates whether all modules are jointly optimized under a unified objective.}

  \label{table:landscape_of_different_audio_tokenizers}
\resizebox{\textwidth}{!}{
\begin{tabular}{@{} c c c c c c c c c c c c @{}}
  \toprule
    \multirow{2}{*}{\textbf{Model}}
    & \multirow{2}{*}{\makecell[c]{\textbf{Frame}\\\textbf{rate}}}
    & \multirow{2}{*}{\makecell[c]{\textbf{Encoder}\\\textbf{Arch.}}}
    & \multirow{2}{*}{\makecell[c]{\textbf{Decoder}\\\textbf{Arch.}}}
    & \multirow{2}{*}{\makecell[c]{\textbf{Stream--}\\\textbf{ing}}}
    & \multirow{2}{*}{\makecell[c]{\textbf{Variable}\\\textbf{Bitrate}}}
    & \multirow{2}{*}{\makecell[c]{\textbf{Semantic}\\\textbf{rich}}}
    & \multicolumn{3}{c}{\textbf{Reconstruction}}
    & \multirow{2}{*}{\makecell[c]{\textbf{Pretrained}\\\textbf{encoder free}}}
    & \multirow{2}{*}{\makecell[c]{\textbf{End-to-end}\\\textbf{optimize}}} \\
  \cmidrule(lr){8-10}
    & & & & & & &
    \textbf{Speech} & \textbf{Sound} & \textbf{Music}
    & & \\
  \midrule

  \textbf{Encodec}               & 75   & CNN         & CNN         & \cmark & \cmark & \xmark & \cmark & \cmark & \cmark & \cmark & \cmark \\
  \textbf{DAC}                   & 75   & CNN         & CNN         & \xmark & \cmark & \xmark & \cmark & \cmark & \cmark & \cmark & \cmark \\
  \textbf{SpeechTokenizer}       & 50   & CNN         & CNN         & \xmark & \cmark & \cmark & \cmark & \xmark & \xmark & \xmark & \xmark \\
  \textbf{Mimi}                  & 12.5 & Hybrid      & Hybrid      & \cmark & \cmark & \cmark & \cmark & \cmark & \cmark & \xmark & \xmark \\
  \textbf{BigCodec}              & 80   & CNN         & CNN         & \xmark & \xmark & \xmark & \cmark & \xmark & \xmark & \cmark & \cmark \\
  \textbf{StableCodec}          & 25   & Hybrid      & Hybrid      & \xmark & \cmark & \xmark & \cmark & \xmark & \xmark & \cmark & \cmark \\
  \textbf{XCodec2.0}             & 50   & Hybrid      & Hybrid      & \xmark & \xmark & \cmark & \cmark & \xmark & \xmark & \xmark & \xmark \\
  \textbf{XY-Tokenizer}          & 12.5 & Hybrid      & Hybrid      & \xmark & \xmark & \cmark & \cmark & \xmark & \xmark & \xmark & \xmark \\
  \textbf{DualCodec}             & 12.5 & Hybrid      & CNN         & \xmark & \cmark & \cmark & \cmark & \xmark & \xmark & \xmark & \xmark \\
  \textbf{Higgs-Audio-Tokenizer} & 25   & Hybrid      & CNN         & \xmark & --    & \cmark & \cmark & \cmark & \cmark & \xmark & \xmark \\
  \textbf{MiMo-Audio-Tokenizer}  & 25   & Hybrid      & Hybrid      & \xmark & \cmark & \cmark & \cmark & \cmark & \cmark & \cmark & \xmark \\
  \textbf{Qwen3-TTS-Tokenizer}   & 12.5 & Hybrid      & Hybrid      & \cmark & --    & \cmark & \cmark & -- & -- & \xmark    & \xmark    \\
  \rowcolor{oursgray}
  \textbf{MOSS-Audio-Tokenizer}            & 12.5 & Trans. & Trans. & \cmark & \cmark & \cmark & \cmark & \cmark & \cmark & \cmark & \cmark \\

  \bottomrule
\end{tabular}
}

\end{table*}

\begin{wrapfigure}{r}{0.45 \textwidth}
  \vspace{-40pt}
  \includegraphics[width=0.9\linewidth]{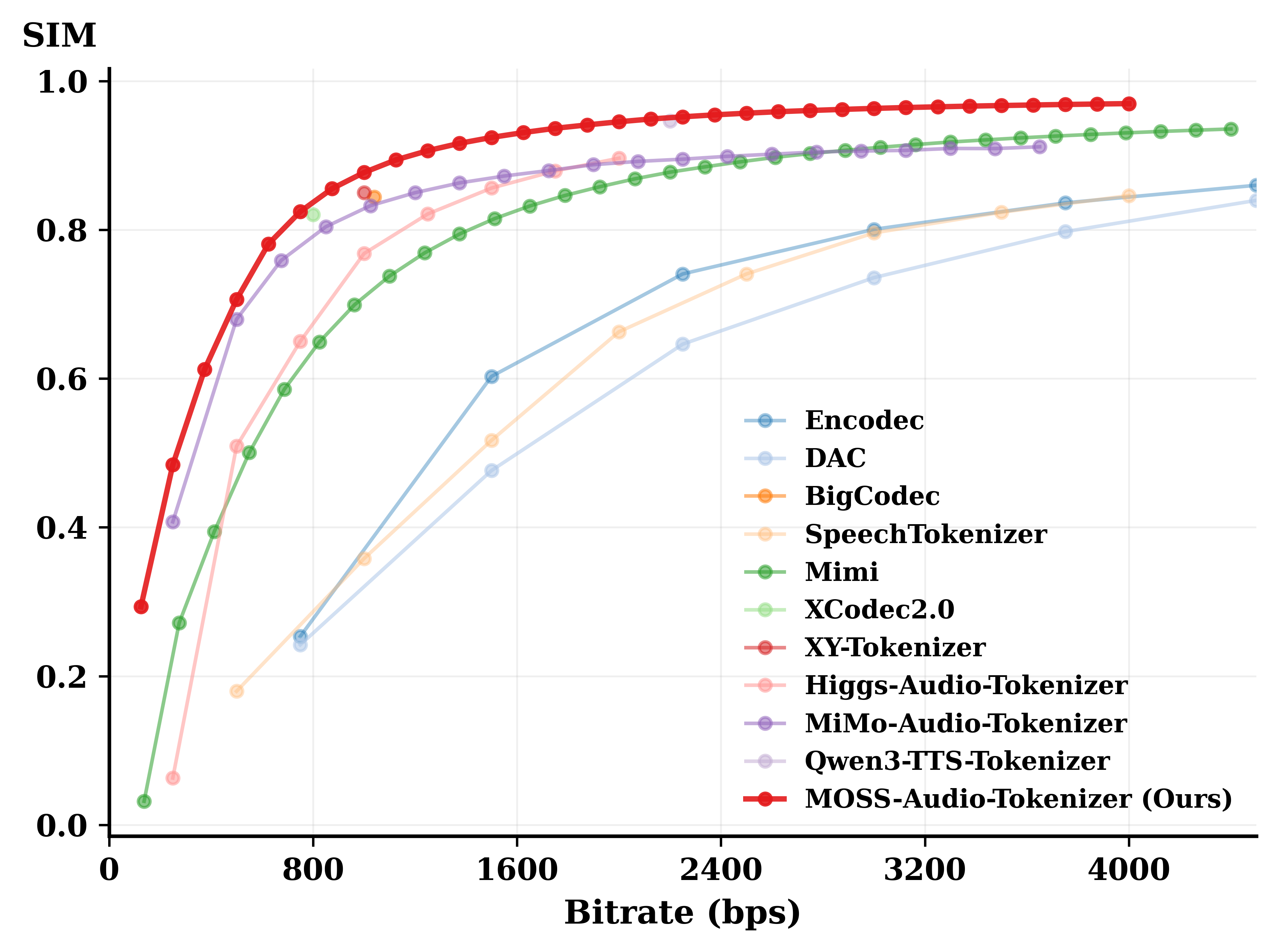}
  \vspace{-10pt}
  \caption{Audio reconstruction quality comparison. }\label{fig:all_tokenizers_sim_vs_bps}
  % between MOSS-Audio-Tokenizer and open-source audio tokenizers.}
  % \label{fig:all_tokenizers_sim_vs_bps}
  \vspace{-30pt}
\end{wrapfigure}

% 问题是什么: 需要一个原生 audio 的 unified tokenizer
Recent advances in large language models~\citep{brown2020language,touvron2023llama,achiam2023gpt,hurst2024gpt,dubey2024llama,yang2024qwen2,yang2025qwen3,guo2025deepseek} have demonstrated the effectiveness of autoregressive modeling over discrete token sequences. By providing a unified discrete interface, text tokenizers~\citep{sennrich2016neural,kudo2018sentencepiece} allow large language models to operate directly on raw text, serving as the foundation upon which compression, understanding, generation, and in-context learning capabilities emerge within a single autoregressive modeling framework. Extending this paradigm to audio requires a unified discrete audio tokenizer that can serve as a native interface for autoregressive modeling~\citep{borsos2023audiolm,agostinelli2023musiclm,yang2024uniaudio,zhang2025mimo}.

% \begin{wrapfigure}{r}{0.42 \textwidth}
% % \vspace{-30pt}
%   \includegraphics[width=\linewidth]{Images/all_tokenizers_sim_vs_bps_v2.png}
%   % \vspace{-20pt}
%   \caption{Audio reconstruction quality comparison between MOSS-Audio-Tokenizer and open-source audio tokenizers.}
%   \label{fig:all_tokenizers_sim_vs_bps}
%   % \vspace{-55pt}
% \end{wrapfigure}

% 相关工作 && 相关工作的不足和转机
Unlike text, audio contains both fine-grained acoustic details and long-range structure, making discrete tokenization more challenging~\citep{borsos2023audiolm}. A unified audio tokenizer should enable high-fidelity reconstruction of diverse audio signals while remaining compatible with autoregressive sequence modeling~\citep{defossez2024moshi,li2025baichuan,zhang2025mimo}. Existing approaches typically address these requirements through pretrained encoders~\citep{hsu2021hubert,chung2021w2v,radford2023robust,ye2025codec,gong2025xy,li2025dualcodec}, multi-stage training pipelines~\citep{wu2023audiodec,welker2025flowdec,zhang2025mimo}, or architecture-specific inductive biases~\citep{zeghidour2021soundstream,defossez2022high,kumar2023high}, achieving strong performance under particular design choices. However, such designs introduce additional dependencies and architectural constraints that make it difficult to scale models, data, and quantization capacity in a unified manner. From this perspective, we draw inspiration from the success of large language models, where simple and efficient architectures trained on large-scale data have proven critical for achieving strong performance~\citep{kaplan2020scaling, henighan2020scaling}. 
We posit that enabling an audio tokenizer to reach a higher performance ceiling similarly requires a simple and scalable architecture that can be trained end-to-end on large amounts of data. Such a design emphasizes joint optimization and scale, while minimizing reliance on external priors, pretrained components, or complex architectural heuristics.

% \zyf{TODO}
In this work, we propose MOSS-Audio-Tokenizer, a fully end-to-end audio tokenizer that serves as a unified discrete interface for autoregressive audio language models. Our tokenizer, build on \textbf{CAT} (\textbf{C}ausal \textbf{A}udio \textbf{T}okenizer with \textbf{T}ransformer)  architecture, operates at a 24\,kHz sampling rate with a low token frame rate of 12.5\,Hz, and jointly optimizes the encoder, quantizer, decoder, and discriminator within a single training pipeline, without relying on pretrained encoders, distillation, or separate optimization of individual components. Both the encoder and decoder are built entirely from causal Transformer blocks, resulting in a simple and scalable architecture that is naturally aligned with autoregressive modeling~\citep{vaswani2017attention}. All components of MOSS-Audio-Tokenizer are designed to operate in a streaming manner, enabling low-latency inference and consistent training–inference behavior~\citep{zeghidour2021soundstream,defossez2022high,defossez2024moshi}.

By scaling large amounts of paired audio–text data, MOSS-Audio-Tokenizer learns discrete representations that are both structurally rich and acoustically expressive, while remaining robust across a wide range of bitrates. As a result, the tokenizer achieves high-quality reconstruction of general audio, including speech, sound, and music, from very low to high bitrate regimes, providing a strong lower bound and a high upper bound for downstream audio language models.

Across speech, sound, and music, MOSS-Audio-Tokenizer achieves state-of-the-art reconstruction quality at all evaluated bitrates. Leveraging its discrete tokens, we further introduce a purely autoregressive text-to-speech model with a Progressive Sequence Dropout training strategy, which naturally exploits the tokenizer's robustness across bitrates and, for the first time, enables a fully autoregressive discrete TTS system~\citep{liao2024fish,ye2025llasa,wang2025spark,xie2025fireredtts} to outperform prior non-autoregressive~\citep{eskimez2024e2,chen2025f5} and cascaded approaches~\citep{betker2023better,wang2023neural,anastassiou2024seed,du2024cosyvoice,wang2024maskgct,zhang2025minimax,zhou2025indextts2,cui2025glm}. In addition, MOSS-Audio-Tokenizer supports competitive automatic speech recognition performance without requiring an auxiliary audio encoder, matching or exceeding models that rely on dedicated audio encoders combined with large language models~\citep{chu2024qwen2,liu2025voxtral,xu2025fireredasr,qwen2.5omni,qwen3omni}. Together, these results demonstrate that CAT architecture provides a scalable and effective foundation for audio compression, understanding, and generation within a unified autoregressive framework.

Our contributions can be summarized as follows:
\begin{itemize}

% [架构：强调是架构名，强调简单易 scale]
\item \textbf{Homogeneous and Scalable Architecture:} We propose \textbf{CAT} (\textbf{C}ausal \textbf{A}udio \textbf{T}okenizer with \textbf{T}ransformer), a purely Transformer-based architecture for discrete audio tokenization. By utilizing a homogeneous stack of causal Transformer blocks, CAT provides a simple and highly scalable discrete interface, minimizing fixed inductive biases and facilitating effective model scaling.

% [规模与通用性：明确参数、数据、比特率范围、流式]
\item \textbf{Large-Scale General Tokenization:} Based on the CAT architecture, we develop \textbf{MOSS-Audio-Tokenizer}, a 1.6-billion-parameter audio tokenizer pre-trained from scratch on 3 million hours of diverse audio. It achieves high-fidelity reconstruction across speech, sound, and music at a low frame rate of 12.5\,Hz. The model natively supports variable bitrates ranging from 0.125\,kbps to 4\,kbps and enables low-latency, frame-level streaming encoding and decoding for real-time applications.

% [下游任务突破：AR TTS 的里程碑和新策略]
\item \textbf{Breakthrough in End-to-End Autoregressive Audio Generation:} Leveraging CAT's discrete tokens, we develop the first purely autoregressive (AR) TTS system that outperforms prior non-autoregressive and cascaded models. Furthermore, we propose \textbf{Progressive Sequence Dropout}, a training strategy that enables a single autoregressive model to perform variable-bitrate speech generation by effectively utilizing the tokenizer's hierarchical quantization structure.

% [分析：强调计算量和参数量增加带来的一致性提升]
\item \textbf{Consistent Scaling Performance:} We investigate the scaling behavior of the CAT architecture with respect to model parameters and training computation (via batch size). Our results demonstrate that CAT exhibits consistent performance gains in reconstruction quality as the model capacity and total computational budget increase, establishing it as a unified and robust foundation for future large-scale audio foundation models.

\end{itemize}

\section{Rethinking Discrete Audio Tokenization for Future Audio Foundation Models}

We rethink discrete audio tokenization from the perspective of autoregressive audio language modeling. Analogous to text tokenizers in large language models, a discrete audio tokenizer should serve as a native interface that bridges raw audio signals and autoregressive sequence modeling~\citep{zhang2023speechtokenizer,defossez2024moshi,zhang2025mimo}. This viewpoint places stringent requirements on the structure, representation capacity, and scalability of the tokenizer, beyond traditional objectives of audio compression or reconstruction.

From this perspective, we identify several key design principles that an audio tokenizer must satisfy in order to effectively support autoregressive audio language models.

\paragraph{Unified Audio Representation.}
A tokenizer should provide a unified discrete representation capable of modeling and reconstructing diverse audio domains, including speech, sound, and music.
Crucially, the resulting tokens should preserve both fine-grained acoustic information and semantic structure, enabling them to function as a meaningful sequence for autoregressive modeling rather than merely a compressed signal.

\paragraph{Simplicity and Scalability.}
To enable efficient scaling with model capacity, data, and computation, the tokenizer architecture should remain simple and homogeneous. Excessive architectural heterogeneity or reliance on specialized components can introduce fixed bottlenecks that hinder joint scaling and limit the effectiveness of large-scale training.

\paragraph{Causality.} 
For compatibility with autoregressive generation and low-latency inference, tokenization should be strictly causal, ensuring that each token is computed without access to future audio context. This property aligns the tokenizer with the operational constraints of autoregressive audio language models and avoids discrepancies between training and inference.

\paragraph{Low Frame Rate and Bitrate Robustness.}
An effective audio tokenizer should operate at a low frame rate to reduce downstream sequence modeling complexity, while remaining robust across a wide range of bitrates. Such flexibility allows a single tokenizer to support diverse downstream tasks, including audio reconstruction, understanding, and generation, without requiring task-specific redesign.

% Guided by these design principles, we instantiate them in \textbf{CAT}, a fully end-to-end, Transformer-based audio tokenizer. CAT jointly optimizes the encoder, quantizer, and decoder within a unified  framework, providing a concrete realization of the above principles. 

\section{Causal Audio Tokenizer with Transformer (CAT)}
\label{main:method}

\begin{figure*}[t!]
  \centering
  \includegraphics[width=0.8\linewidth]{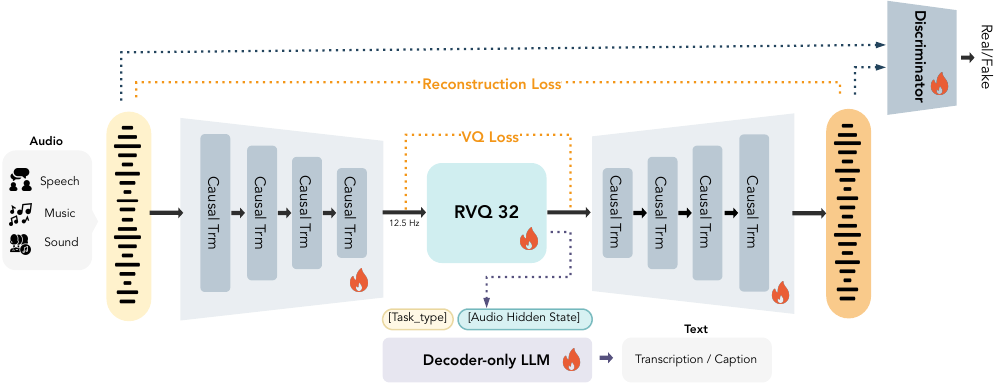}
  \caption{Architecture of \textbf{CAT} (\textbf{C}ausal \textbf{A}udio \textbf{T}okenizer with \textbf{T}ransformer).
Both the encoder and decoder are built upon causal Transformers.
All components, including the encoder, quantizer, decoder, causal language model, and discriminator, are optimized jointly in an end-to-end manner.}
  \label{fig:TAC_arch}
\end{figure*}

\subsection{Homogeneous Architecture for Scalable Audio Tokenization}
A central design goal of CAT is to enable scalable audio tokenization that can seamlessly integrate with large autoregressive and multimodal foundation models.
To this end, we adopt a \textit{CNN-free} architecture that is built entirely upon causal Transformer blocks, as illustrated in Figure~\ref{fig:TAC_arch}.
Compared to prior neural audio codecs that rely heavily on convolutional inductive biases or hybrid CNN--Transformer designs, our approach deliberately minimizes architectural specialization, favoring simplicity, uniformity, and scalability.

\paragraph{Fully Transformer-based encoder--decoder.}

Both the encoder and decoder in CAT are implemented as stacks of causal Transformer blocks, 
forming a CNN-free architecture and enabling streaming encoding and decoding.
CAT operates directly on raw audio waveforms at both the input and output, avoiding intermediate signal representations such as mel-spectrograms.
The input waveform is first \textit{patchified} into a sequence of fixed-dimensional vectors and processed by the causal Transformer encoder.
To progressively compress long audio sequences into a compact representation, we insert patchify operations between Transformer blocks, which gradually reduce the temporal resolution.
As a result, the encoder maps 24\,kHz waveforms into discrete token sequences at an average rate of 12.5 frames per second.
The decoder mirrors this process in reverse, reconstructing the waveform from discrete tokens in a fully causal manner. Further implementation details are provided in Appendix~\ref{appendix:implementation_details}.

\paragraph{Scalable residual vector quantization.}
For discretization, we employ residual vector quantization (RVQ).
To support robust modeling across a wide range of bitrates, we adopt $N_q = 32$ residual quantization layers and enable quantizer dropout during training.
This variable-bitrate design directly facilitates the controllable audio generation framework introduced later.

\subsection{Unified Audio Modeling}
We use multi-task learning to enable CAT to achieve both strong alignment with text and high-quality audio reconstruction.
\paragraph{Semantic Modeling via Audio-to-Text Tasks.}
To encourage the token representation to be semantically rich and aligned with text-based language modeling, we incorporate an auxiliary audio-to-text objective. Specifically, we employ a 0.5B-parameter decoder-only LLM~\citep{qwen2.5} and condition it on the representations produced by CAT. Concretely, we feed the hidden states from the quantizer output into the LLM, which then autoregressively predicts textual tokens. We consider a diverse set of audio-to-text tasks, including automatic speech recognition (ASR), multi-speaker ASR, and audio captioning. For audio samples that are paired with textual annotations, we apply the corresponding semantic modeling objective. Each task is specified by a fixed task tag \( \mathcal{T} \), which is prepended to the LLM input. The semantic objective is optimized using a standard cross-entropy loss:
% \begin{equation}
% \label{eq:audio_to_text_loss}
% \mathcal{L}_{\mathrm{sem}}
% =
% \mathbb{E}_{(\mathbf{q}, \mathbf{s}) \sim \mathcal{D}}
% \left[
% - \sum_{t=1}^{|\mathbf{s}|}
% \log p_{\theta_{\mathrm{LLM}}}
% \left(
% \mathrm{s}_t \,\middle|\,
% \mathcal{T},\, \mathbf{q},\, \mathrm{s}_{<t}
% \right)
% \right],
% \end{equation}
\begin{equation}
\label{eq:audio_to_text_loss}
\mathcal{L}_{\mathrm{sem}}
=
- \sum_{t=1}^{|\mathrm{s}|}
\log p_{\theta_{\mathrm{LLM}}}
\left(
\mathrm{s}_t \,\middle|\,
\mathcal{T},\, \mathbf{q},\, \mathrm{s}_{<t}
\right),
\end{equation}
where $\mathbf{s} = (\mathrm{s}_1, \dots, \mathrm{s}_{|\mathbf{s}|})$
denotes the target text token sequence,
$\mathbf{q}$ denotes the sequence of quantized audio representations produced by CAT,
$\mathcal{T}$ is a task-specific prompt token,
and $\theta_{\mathrm{LLM}}$ are the parameters of the causal language model.

% where $\mathbf{s} = (\mathrm{s}_1, \dots, \mathrm{s}_{|\mathbf{s}|})$
% denotes the target text token sequence.
% $\mathbf{q}$ denotes the sequence of quantized audio representations produced by CAT,
% $\mathcal{T}$ is a task-specific prompt token,
% $\mathcal{D}$ is the set of paired audio--text samples,
% and $\theta_{\mathrm{LLM}}$ are the parameters of the causal language model.

% \begin{equation}
% \label{eq:audio_to_text_loss}
% \mathcal{L}_{\mathrm{sem}}
% =
% \mathbb{E}_{(\mathbf{q}, \mathbf{s}) \sim \mathcal{D}}
% \left[
% - \sum_{t=1}^{|\mathbf{s}|}
% \log p_{\theta_{\mathrm{LLM}}}
% \left(
% \mathbf{s}_t \,\middle|\,
% \mathcal{T},\, \mathbf{q},\, \mathbf{s}_{<t}
% \right)
% \right],
% \end{equation}
% where $\mathbf{s} = (s_1, \dots, s_{|\mathbf{s}|})$ denotes the target text token sequence,
% $s_{<t}$ represents the preceding text tokens,
% $\mathbf{q}$ denotes the sequence of quantized audio representations produced by CAT,
% $\mathcal{T}$ is a task-specific prompt token,
% $\mathcal{D}$ is the set of paired audio--text samples,
% and $\theta_{\mathrm{LLM}}$ are the parameters of the causal language model.

% \begin{equation}
% \label{eq:audio_to_text_loss}
% \mathcal{L}_{\mathrm{sem}} =
% - \sum_{i=1}^{N}
% \log p_{\theta_{\mathrm{LLM}}}
% \!\left(
% s_i \mid \mathcal{T},\, q,\, s_{<i}
% \right),
% \end{equation}
% where \( \{s_i\}_{i=1}^{N} \) denotes the target text token sequence, \( s_{<i} \) represents the preceding text tokens, \( q \) denotes the LLM input derived from the audio hidden states, and \( \theta_{\mathrm{LLM}} \) are the parameters of the causal language model.

\paragraph{Quantizer Optimization.}
For training simplicity and stability, each quantization layer in CAT adopts
factorized vector quantization~\citep{kumar2023high},
where codebooks are directly optimized via gradient descent,
without relying on additional codebook update mechanisms~\citep{defossez2022high}. We incorporate a commitment loss and a codebook loss to jointly
optimize the encoder and the codebook entries:
\begin{equation}
\label{eq:cmt_loss}
\mathcal{L}_{\mathrm{cmt}} = \sum_{c=1}^{N_q}
\left\| \mathbf{z}_c - \operatorname{sg}(q_c(\mathbf{z}_c)) \right\|_2^2,
\end{equation}
\begin{equation}
\label{eq:codebook_loss}
\mathcal{L}_{\mathrm{code}} = \sum_{c=1}^{N_q}
\left\| \operatorname{sg}(\mathbf{z}_c) - q_c(\mathbf{z}_c) \right\|_2^2,
\end{equation}
where \( \mathbf{z}_c \) denotes the input to the \( c \)-th quantization layer,
\( q_c(\mathbf{z}_c) \) is the corresponding quantized output,
\( N_q \) is the number of quantizers,
and \( \operatorname{sg}(\cdot) \) denotes the stop-gradient operator~\citep{van2017neural}.

\paragraph{Acoustic Modeling via Reconstruction Tasks.}
To ensure high-fidelity and domain-robust audio reconstruction, we adopt a multi-scale mel-spectrogram loss:
\begin{equation}
\label{eq:rec_loss}
\mathcal{L}_{\mathrm{rec}} =
\sum_{i=5}^{11}
\left\|
S_{2^i}(\mathbf{x}) - S_{2^i}(\hat{\mathbf{x}})
\right\|_1,
\end{equation}
where \( S_{2^i}(\cdot) \) denotes the mel-spectrogram computed using a normalized short-time Fourier transform (STFT) with window size \( 2^i \) and hop size \( 2^{i-2} \). Here, \( \mathbf{x} \) is the ground-truth waveform and \( \hat{\mathbf{x}} \) is the reconstructed waveform generated by the decoder.

% \paragraph{Adversarial Training.}
% To further improve reconstruction fidelity and perceptual quality, we employ adversarial training with multiple discriminators, which help capture fine-grained temporal and spectral characteristics of audio signals.

% The discriminator loss is defined as:
% \begin{equation}
% \label{eq:disc_loss}
% \mathcal{L}_{D}(\mathbf{x}, \hat{\mathbf{x}}) =
% \frac{1}{K}
% \sum_{k=1}^{K}
% \left[
% (1 - D_k(\mathbf{x}))^2 + D_k^2(\hat{\mathbf{x}})
% \right],
% \end{equation}
% where \( D_k \) denotes the \( k \)-th discriminator and \( K \) is the total number of discriminators. The corresponding adversarial loss for the generator is given by:
% \begin{equation}
% \label{eq:adv_loss}
% \mathcal{L}_{\mathrm{adv}}(\hat{\mathbf{x}}) =
% \frac{1}{K}
% \sum_{k=1}^{K}
% (1 - D_k(\hat{\mathbf{x}}))^2.
% \end{equation}

% We also incorporate a feature matching loss:
% \begin{equation}
% \label{eq:feature_matching_loss}
% \mathcal{L}_{\mathrm{feat}}(\mathbf{x}, \hat{\mathbf{x}}) =
% \frac{1}{K}
% \sum_{k=1}^{K}
% \frac{1}{L_k}
% \sum_{l=1}^{L_k}
% \frac{
% \left\|
% D_k^l(\mathbf{x}) - D_k^l(\hat{\mathbf{x}})
% \right\|_1
% }{
% \operatorname{mean}\!\left(
% \left\| D_k^l(\mathbf{x}) \right\|_1
% \right)
% },
% \end{equation}
% where \( D_k^l(\cdot) \) denotes the feature representation from the \( l \)-th layer of the \( k \)-th discriminator, and \( L_k \) is the number of layers in discriminator \( D_k \).

\paragraph{Adversarial Training.}
To further improve reconstruction fidelity and perceptual quality, we employ adversarial training with multiple discriminators.
Specifically, we adopt the discriminator architecture and training objectives,
including the adversarial loss, feature matching loss and discriminator loss,
following XY-Tokenizer~\citep{gong2025xy}.

\paragraph{Overall Training Objective.}
The overall generator objective is a weighted combination of all loss terms:
\begin{equation}
\label{eq:generator_loss}
\begin{aligned}
\mathcal{L}_{\mathrm{G}} =\;&
\lambda_{\mathrm{sem}} \mathcal{L}_{\mathrm{sem}}
+ \lambda_{\mathrm{rec}} \mathcal{L}_{\mathrm{rec}}
+ \lambda_{\mathrm{cmt}} \mathcal{L}_{\mathrm{cmt}} 
+ \lambda_{\mathrm{code}} \mathcal{L}_{\mathrm{code}}
+ \lambda_{\mathrm{adv}} \mathcal{L}_{\mathrm{adv}}
+ \lambda_{\mathrm{feat}} \mathcal{L}_{\mathrm{feat}} ,
\end{aligned}
\end{equation}
where $\mathcal{L}_{\mathrm{adv}}$ and $\mathcal{L}_{\mathrm{feat}}$
denote the adversarial and feature matching losses defined in
XY-Tokenizer~\citep{gong2025xy}. \( \lambda_{\mathrm{sem}} \), \( \lambda_{\mathrm{rec}} \), \( \lambda_{\mathrm{cmt}} \), \( \lambda_{\mathrm{code}} \), \( \lambda_{\mathrm{adv}} \), \( \lambda_{\mathrm{feat}} \) are scalar hyperparameters controlling the relative contribution of each loss term.

% \paragraph{Overall Training Objective.}

All components of CAT, including the encoder, quantizer, decoder, and discriminators, are optimized jointly in an end-to-end manner. By scaling large amounts of audio data, CAT learns to achieve both high-fidelity reconstruction of general audio and semantically rich discrete representations, without relying on pretrained encoders or external semantic teachers~\citep{hsu2021hubert,chung2021w2v,zhang2023speechtokenizer,ye2025codec}.

% \zyf{TODO: semantic modeling, reconstruction object, adversarial training, and overall training object}

\subsection{Bitrate Controllable Audio Generation}

% \begin{wrapfigure}{r}{0.50\textwidth}
% \vspace{-14pt}
%   \centering
%   \includegraphics[width=\linewidth]{Images/TTS_arch.pdf}
% \caption{
% Architecture of bitrate controllable audio modeling.
% During training, Progressive Sequence Dropout randomly truncates the number of active RVQ layers.
% During inference, when decoding with a fixed depth $k$, only the first $k$ RVQ tokens are provided as input at each time step, and the Depth Transformer autoregressively predicts only these $k$ tokens, while finer RVQ layers are omitted.
% }
%   \label{fig:tts_arch}
% \vspace{-14pt}
% \end{wrapfigure}

% \begin{figure}[t]
% \vspace{-7pt}
%   \centering
%   \includegraphics[width=0.9\linewidth]{Images/TTS_arch.pdf}
% \caption{
% Speech Generation architecture with Progressive Sequence Dropout.
% During training, Progressive Sequence Dropout randomly truncates the number of active RVQ layers.
% During inference, when decoding with a fixed depth $k$, only the first $k$ RVQ tokens are provided as input at each time step, and the Depth Transformer autoregressively predicts only these $k$ tokens, while finer RVQ layers are omitted.
% }
%   \label{fig:tts_arch}
% \vspace{-10pt}
% \end{figure}

\noindent\textbf{End-to-end variable-bitrate autoregressive speech generation.}
Building on CAT, we construct \textbf{CAT-TTS}, a fully end-to-end, purely autoregressive speech generation model that supports variable-bitrate synthesis.
The model directly generates speech from text tokens and a speaker prompt by predicting CAT's RVQ tokens at a controllable depth, without requiring semantic disentanglement~\citep{zhang2023speechtokenizer,zhang2024speechgpt,zhang2025vevo} or cascading multiple generative models~\citep{du2024cosyvoice,anastassiou2024seed,cui2025glm}.
By leveraging CAT as a unified discrete interface, both linguistic content and acoustic information are modeled within a single autoregressive framework.

\noindent\textbf{Autoregressive modeling over RVQ tokens.}
Since CAT represents audio using residual vector quantization (RVQ), we adopt the \textbf{Temporal Transformer + Depth Transformer} architecture~\citep{defossez2024moshi} for multi-stream autoregressive modeling.
The Temporal Transformer captures long-range dependencies along the temporal dimension, while the Depth Transformer models the coarse-to-fine residual structure across RVQ layers within each time step.
Under this design, each RVQ token conditions only on tokens from previous time steps and on preceding RVQ layers at the current time step, ensuring strict causality without information leakage.

\noindent\textbf{Progressive Sequence Dropout.}
To enable robust generation across a wide range of bitrates within a single model, we propose \textit{Progressive Sequence Dropout}, a simple yet effective training strategy that requires \textbf{no architectural modifications or additional parameters}.
During training, dropout is activated with probability $p$.
When activated, we uniformly sample a prefix length $K \in \{1, \dots, N_q - 1\}$, where $N_q$ denotes the total number of RVQ layers, and discard RVQ tokens from layers $K{+}1$ to $N_q$.
Otherwise, all RVQ layers are retained.
This strategy exposes the model to truncated RVQ prefixes during training, where a prefix length is randomly sampled independently for each training sample, encouraging the model to learn conditional generation under varying bitrates.

\paragraph{Prefix definition.}
We introduce a Bernoulli random variable
\begin{equation}
z \sim \mathrm{Bernoulli}(p),
\end{equation}
where $z = 1$ indicates that Progressive Sequence Dropout is applied and $z = 0$ otherwise.
When $z = 1$, the prefix length $K$ is sampled uniformly as described above; when $z = 0$, we set $K = N_q$.
The effective number of active RVQ layers is then defined as
\begin{equation}
\hat{K} = (1 - z)\, N_q + z\, K .
\end{equation}

\paragraph{Global input aggregation and training objective.}
Let $\mathbf{q}_{t,k}$ denote the RVQ token at time step $t$ and layer $k$, and let $\mathrm{Emb}_k(\cdot)$ denote the embedding lookup table for the $k$-th RVQ codebook.
For each time step $t$, the speech input to the Temporal Transformer is constructed by aggregating the embeddings of the first $\hat{K}$ RVQ layers:
\begin{equation}
\tilde{\mathbf{e}}_t
=
\sum_{k=1}^{\hat{K}}
\mathrm{Emb}_k \bigl( \mathbf{q}_{t,k} \bigr) .
\end{equation}
The Temporal Transformer processes the resulting acoustic embedding sequence
$\{ \tilde{\mathbf{e}}_t \}_{t=1}^{T}$
using a causal attention mask along the temporal dimension.

% \paragraph{Training objective.}
The Depth Transformer predicts RVQ tokens autoregressively along the depth dimension.
The training loss is computed only over the retained RVQ prefix:
\begin{equation}
\mathcal{L}
=
-
\sum_{t=1}^{T}
\sum_{k=1}^{\hat{K}}
\log
p_\theta
\Bigl(
\mathbf{q}_{t,k}
\mid
\mathbf{x},\,
\mathbf{q}_{<t},\,
\mathbf{q}_{t,<k}
\Bigr) ,
\end{equation}
where $\theta$ denotes the model parameters, $\mathbf{x}$ represents the input text token sequence, $\mathbf{q}_{<t}$ denotes all RVQ tokens from previous time steps, and $\mathbf{q}_{t,<k}$ denotes RVQ tokens from preceding layers at the same time step.

\paragraph{Inference.}
At inference time, we explicitly control the synthesis bitrate by selecting an inference depth $K_{\mathrm{infer}}$.
The Temporal Transformer takes as input the text tokens together with the first $K_{\mathrm{infer}}$ RVQ token streams at each time step.
The Depth Transformer then autoregressively predicts only these $K_{\mathrm{infer}}$ RVQ layers, while finer layers are omitted.
Finally, the predicted RVQ tokens from the first $K_{\mathrm{infer}}$ layers are decoded into waveforms using the CAT decoder.
As CAT is trained with quantizer dropout, the decoder is inherently robust to varying effective bitrates, which aligns naturally with Progressive Sequence Dropout in the speech generation model.
% Finally, the predicted RVQ tokens are decoded into waveforms using the CAT decoder. 

\paragraph{Special case.}
When $p = 0$, Progressive Sequence Dropout is disabled, yielding $z = 0$ for all training samples and $\hat{K} = N_q$.
In this case, the proposed method reduces exactly to the standard Temporal Transformer + Depth Transformer formulation for multi-stream autoregressive speech generation.

% \paragraph{Special case.}
% When $p = 0$, Progressive Sequence Dropout is disabled ($z=0$ for all samples) and $\hat{K} = N_q$.
% In this case, the proposed method reduces exactly to the standard Temporal Transformer + Depth Transformer formulation for multi-sequence autoregressive speech generation.

% \newcommand{\up}{$\uparrow$}
% \newcommand{\down}{$\downarrow$}
% \newcommand{\bnum}[1]{\textbf{#1}}
% \newcolumntype{Y}{>{\centering\arraybackslash}X}

\begin{table*}[t]
  \centering
  \fontsize{8pt}{10pt}\selectfont
  \setlength{\tabcolsep}{4pt}
  \renewcommand{\arraystretch}{1.15}
  \caption{Reconstruction quality comparison of open-source audio tokenizers on speech and audio/music data.
Speech metrics are evaluated on LibriSpeech test-clean (English) and AISHELL-2 (Chinese) and reported as English/Chinese.
Audio metrics are evaluated on the AudioSet evaluation subset, while music metrics are evaluated on the MUSDB dataset; values are reported as audio/music.
STFT-Dist. denotes the STFT distance.
Higher is better for speech metrics, whereas lower is better for audio/music metrics. $\boldsymbol{N}_{\mathrm{VQ}}$ denotes the number of quantizers.}
  \label{table:codec_recon_all}

  % 4 attribute cols + 4 speech metrics + 2 audio/music metrics = 10 cols
  \begin{tabular}{@{} c c c c c c c c c c @{}}
\toprule
\multirow{2}{*}{\textbf{Model}}
& \multirow{2}{*}{\textbf{bps}}
& \multirow{2}{*}{\makecell[c]{\textbf{Frame}\\\textbf{rate}}}
& \multirow{2}{*}{$\boldsymbol{N}_{\mathrm{VQ}}$}
& \multicolumn{4}{c}{\textbf{Speech}}
& \multicolumn{2}{c}{\textbf{Audio / Music}} \\
\cmidrule(lr){5-8}\cmidrule(lr){9-10}
& & & 
& \textbf{SIM}\,\up
& \textbf{STOI}\,\up
& \textbf{PESQ-NB}\,\up
& \textbf{PESQ-WB}\,\up
& \textbf{Mel-Loss}\,\down
& \textbf{STFT-Dist.}\,\down \\
\midrule

\textbf{StableCodec} & 700  & 25 & 2  & 0.62 / 0.45 & 0.91 / 0.86 & 2.91 / 2.50 & 2.24 / 1.93 & -- / -- & -- / -- \\
\textbf{XCodec2.0} & 800  & 50 & 1  & 0.82 / 0.74 & 0.92 / 0.86 & 3.04 / 2.46 & 2.43 / 1.96 & -- / -- & -- / -- \\
\textbf{MiMo-Audio-Tokenizer} & 850  & 25 & 4  & 0.80 / 0.74 & 0.91 / 0.87 & 2.94 / 2.62 & 2.39 / 2.14 & \bnum{0.82} / 0.81 & 2.33 / 2.23 \\
\textbf{Higgs-Audio-Tokenizer} & 1000  & 25 & 4  & 0.77 / 0.68 & 0.83 / 0.82 & 3.03 / 2.61 & 2.48 / 2.14 & 0.83 / \bnum{0.80} & 2.20 / 2.05 \\
\textbf{SpeechTokenizer} & 1000  & 50 & 2  & 0.36 / 0.25 & 0.77 / 0.68 & 1.59 / 1.38 & 1.25 / 1.17 & -- / -- & -- / -- \\
\textbf{XY-Tokenizer} & 1000  & 12.5 & 8  & 0.85 / 0.79 & 0.92 / 0.87 & 3.10 / 2.63 & 2.50 / 2.12 & -- / -- & -- / -- \\
\textbf{BigCodec} & 1040  & 80 & 1  & 0.84 / 0.69 & 0.93 / 0.88 & 3.27 / 2.55 & 2.68 / 2.06 & -- / -- & -- / -- \\
\textbf{Mimi} & 1100  & 12.5 & 8  & 0.74 / 0.59 & 0.91 / 0.85 & 2.80 / 2.24 & 2.25 / 1.78 & 1.24 / 1.19 & 2.62 / 2.49 \\
\rowcolor{oursgray} \textbf{MOSS-Audio-Tokenizer} & 750  & 12.5 & 6  & 0.82 / 0.75 & 0.93 / 0.89 & 3.14 / 2.73 & 2.60 / 2.22 & 0.86 / 0.85 & 2.21 / 2.10 \\
\rowcolor{oursgray} \textbf{MOSS-Audio-Tokenizer} & 1000  & 12.5 & 8  & \bnum{0.88} / \bnum{0.81} & \bnum{0.94} / \bnum{0.91} & \bnum{3.38} / \bnum{2.96} & \bnum{2.87} / \bnum{2.43} & \bnum{0.82} / \bnum{0.80} & \bnum{2.16} / \bnum{2.04} \\
\midrule
\textbf{DAC} & 1500  & 75 & 2  & 0.48 / 0.41 & 0.83 / 0.79 & 1.87 / 1.67 & 1.48 / 1.37 & -- / -- & -- / -- \\
\textbf{Encodec} & 1500  & 75 & 2  & 0.60 / 0.45 & 0.85 / 0.81 & 1.94 / 1.80 & 1.56 / 1.48 & 1.12 / 1.04 & 2.60 / 2.42 \\
\textbf{Higgs-Audio-Tokenizer} & 2000  & 25 & 8  & 0.90 / 0.83 & 0.85 / 0.85 & 3.59 / 3.22 & 3.11 / 2.73 & 0.74 / 0.70 & 2.07 / 1.92 \\
\textbf{SpeechTokenizer} & 2000  & 50 & 4  & 0.66 / 0.50 & 0.88 / 0.80 & 2.38 / 1.79 & 1.92 / 1.49 & -- / -- & -- / -- \\
\textbf{Qwen3-TTS-Tokenizer} & 2200  & 12.5 & 16  & \bnum{0.95} / 0.88 & \bnum{0.96} / 0.93 & 3.66 / 3.10 & 3.19 / 2.62 & -- / -- & -- / -- \\
\textbf{MiMo-Audio-Tokenizer} & 2250  & 25 & 12  & 0.89 / 0.83 & 0.95 / 0.92 & 3.57 / 3.25 & 3.05 / 2.71 & \bnum{0.70} / \bnum{0.68} & 2.21 / 2.10 \\
\textbf{Mimi} & 2475  & 12.5 & 18  & 0.89 / 0.76 & 0.94 / 0.91 & 3.49 / 2.90 & 2.97 / 2.35 & 1.10 / 1.06 & 2.45 / 2.32 \\
\rowcolor{oursgray} \textbf{MOSS-Audio-Tokenizer} & 1500  & 12.5 & 12  & 0.92 / 0.86 & 0.95 / 0.93 & 3.64 / 3.27 & 3.20 / 2.74 & 0.77 / 0.74 & 2.08 / 1.96 \\
\rowcolor{oursgray} \textbf{MOSS-Audio-Tokenizer} & 2000  & 12.5 & 16  & \bnum{0.95} / \bnum{0.89} & \bnum{0.96} / \bnum{0.94} & \bnum{3.78} / \bnum{3.46} & \bnum{3.41} / \bnum{2.96} & 0.73 / 0.70 & \bnum{2.03} / \bnum{1.90} \\
\midrule
\textbf{DAC} & 3000  & 75 & 4  & 0.74 / 0.67 & 0.90 / 0.88 & 2.76 / 2.47 & 2.31 / 2.07 & 0.86 / 0.83 & 2.23 / 2.10 \\
\textbf{MiMo-Audio-Tokenizer} & 3650  & 25 & 20  & 0.91 / 0.85 & 0.95 / 0.93 & 3.73 / 3.44 & 3.25 / 2.89 & 0.66 / 0.65 & 2.17 / 2.06 \\
\textbf{SpeechTokenizer} & 4000  & 50 & 8  & 0.85 / 0.69 & 0.92 / 0.85 & 3.05 / 2.20 & 2.60 / 1.87 & -- / -- & -- / -- \\
\textbf{Mimi} & 4400  & 12.5 & 32  & 0.94 / 0.83 & 0.96 / 0.94 & 3.80 / 3.31 & 3.43 / 2.78 & 1.02 / 0.98 & 2.34 / 2.21 \\
\textbf{Encodec} & 4500  & 75 & 6  & 0.86 / 0.75 & 0.92 / 0.91 & 2.91 / 2.63 & 2.46 / 2.15 & 0.91 / 0.84 & 2.33 / 2.17 \\
\textbf{DAC} & 6000  & 75 & 8  & 0.89 / 0.84 & 0.95 / 0.94 & 3.75 / 3.57 & 3.41 / 3.20 & \bnum{0.65} / \bnum{0.63} & 1.97 / 1.87 \\
\rowcolor{oursgray} \textbf{MOSS-Audio-Tokenizer} & 3000  & 12.5 & 24  & 0.96 / 0.92 & \bnum{0.97} / \bnum{0.96} & 3.90 / 3.64 & 3.61 / 3.20 & 0.69 / 0.66 & 1.98 / 1.84 \\
\rowcolor{oursgray} \textbf{MOSS-Audio-Tokenizer} & 4000  & 12.5 & 32  & \bnum{0.97} / \bnum{0.93} & \bnum{0.97} / \bnum{0.96} & \bnum{3.95} / \bnum{3.71} & \bnum{3.69} / \bnum{3.30} & 0.68 / 0.64 & \bnum{1.96} / \bnum{1.82} \\

    \bottomrule
  \end{tabular}
\end{table*}

\section{Experiments}
\subsection{Implementation Details}

Building upon the CAT architecture, we develop \textbf{MOSS-Audio-Tokenizer}, a large-scale audio tokenizer featuring 1.6 billion parameters. The model utilizes a causal Transformer-based encoder--decoder paired with hierarchical patching, which facilitates efficient streaming audio modeling. 
Discrete representations are learned using a 32-layer residual vector quantizer with quantizer dropout to support variable-bitrate tokenization.
To encourage semantic alignment, we attach a decoder-only causal language model for audio-to-text supervision.
Training is performed on approximately 3M hours of diverse speech, sound, and music data,
using a combination of reconstruction, semantic, and adversarial objectives. All components of MOSS-Audio-Tokenizer, including the encoder, quantizer, decoder, discriminator, decoder-only LLM are optimized jointly in an end-to-end manner. All architectural details, optimization hyperparameters,
and training schedules are provided in Appendix~\ref{appendix:implementation_details}.

\subsection{Reconstruction Evaluation}
We compare MOSS-Audio-Tokenizer with open-source audio tokenizers using both objective and subjective evaluation metrics across low (750--1500\,bps), medium (1500--2500\,bps), and high (2500--6000\,bps) bitrate regimes.
Table~\ref{table:codec_recon_all} summarizes the objective reconstruction results on speech, general audio, and music benchmarks.

Across all evaluated bitrate regimes, MOSS-Audio-Tokenizer achieves strong performance on speech reconstruction, outperforming prior methods at low bitrates and achieving state-of-the-art results at medium and high bitrates.
On audio and music benchmarks, MOSS-Audio-Tokenizer maintains competitive performance across all evaluated bitrates, with reconstruction quality improving as bitrate increases, indicating that the model effectively benefits from increased bitrate and model capacity through joint end-to-end optimization.

Additional details on the compared open-source audio tokenizers, as well as subjective evaluation results are provided in Appendix~\ref{appendix:evaluation_audio_tokenizer}.

\subsection{Speech Generation}
% 这边先空着

\begin{figure}[h]
  \centering
  \includegraphics[width=1\linewidth]{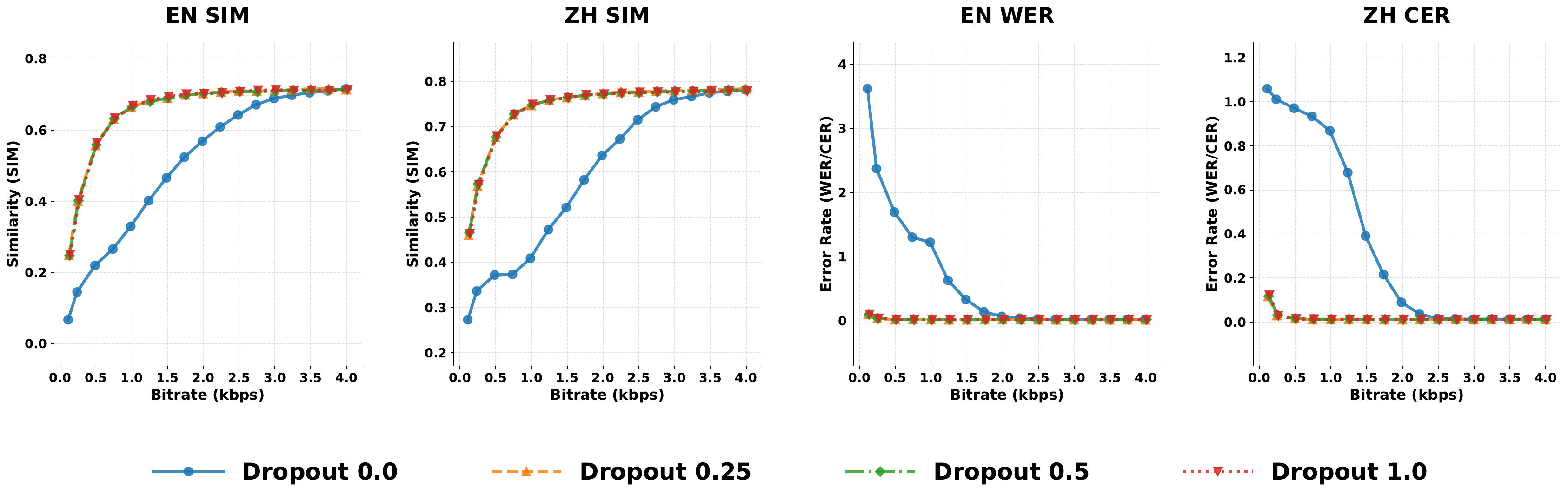}
    \caption{
    Effect of Progressive Sequence Dropout on fully autoregressive TTS across different bitrates.
    }
  \label{fig:TTS_dropout_comparison}
\end{figure}

\paragraph{Experimental Settings.}
We initialize the Temporal Transformer with the pretrained Qwen3-1.7B model~\citep{yang2025qwen3}.
The Depth Transformer consists of four Transformer blocks and is randomly initialized.
We train the model on a mixture of VoxBox, as introduced in SparkAudio~\citep{wang2025spark}, and an internal dataset, totaling approximately 200k hours of speech data.
Evaluation is conducted on the Seed-TTS-Eval benchmark~\citep{anastassiou2024seed}. Training details are provided in Appendix~\ref{appendix:speech_generation}.

\paragraph{Effectiveness of Progressive Sequence Dropout.}
We investigate the effect of Progressive Sequence Dropout by varying the dropout probability $p \in \{0.0, 0.25, 0.5, 1.0\}$, with results summarized in Figure~\ref{fig:TTS_dropout_comparison}.
At full bitrate, all models achieve comparable performance, exhibiting low word error rate and high speaker similarity.
However, as the bitrate decreases, the model trained without dropout exhibits a much steeper degradation in similarity and word error rate.
This stems from the mismatch between training and inference, as the model is trained exclusively with full RVQ depth but evaluated using truncated representations.

In contrast, models trained with Progressive Sequence Dropout are substantially more robust under reduced bitrate settings.
Across different dropout probabilities ($p=0.25, 0.5, \text{ and } 1.0$), TTS performance remains highly consistent at each bitrate, indicating the exact dropout probability has limited impact on generation quality.
Meanwhile, increasing $p$ significantly reduces GPU memory consumption during training.
Therefore, we adopt $p=1.0$ in all subsequent experiments to maximize training efficiency while maintaining comparable synthesis quality.

\definecolor{paradigmgray}{gray}{0.85}

\begin{table}[h]
  \centering
  \fontsize{8pt}{10pt}\selectfont
  \setlength{\tabcolsep}{10.5pt}
  \renewcommand{\arraystretch}{1.15}
  \caption{Comparison with open-source TTS systems on Seed-TTS-Eval.
  Bitrate control indicates whether a TTS system allows explicit specification of the synthesis bitrate at inference time.
  For FlexiCodec-TTS, bitrate is controlled by switching the frame rate of the autoregressive model.
  For CAT-TTS, bitrate is controlled by specifying the number of RVQ tokens generated by the Depth Transformer.}
  \label{table:tts_seed_compare}

    \begin{tabular}{cccccc}
    \toprule
    \multirow{2}{*}{\makecell{\textbf{TTS}\\\textbf{Systems}}} &
    \multirow{2}{*}{\makecell{\textbf{Bitrate}\\\textbf{Control.}}} &
    \multicolumn{2}{c}{\textbf{Seed-EN}} &
    \multicolumn{2}{c}{\textbf{Seed-ZH}} \\
    \cmidrule(lr){3-4} \cmidrule(lr){5-6}
    & & \textbf{WER}$\downarrow$ & \textbf{SIM}$\uparrow$
      & \textbf{CER}$\downarrow$ & \textbf{SIM}$\uparrow$ \\
    \midrule

    \multicolumn{5}{c}{\textbf{Cascade (AR+NAR / NAR+NAR)}} \\
    \midrule
    MaskGCT        & \xmark & 2.62 & 71.7 & 2.27 & 77.4 \\
    FireRedTTS     & \xmark  & 3.84 & 46.0 & 1.51 & 63.5 \\
    CosyVoice2     & \xmark  & 3.09 & 65.9 & 1.38 & 75.7 \\
    Qwen2.5-Omni   & \xmark  & 2.72 & 63.2 & 1.70 & 75.2 \\
    CosyVoice3-1.5B& \xmark  & 2.22 & \bnum{72.0} & 1.12 & \bnum{78.1} \\
    IndexTTS2      & \xmark  & 2.23 & 70.6 & 1.03 & 76.5 \\
    FlexiCodec-TTS & \cmark  & 2.63 & 65.7 & - & - \\
    GLM-TTS        & \xmark  & \bnum{1.91} & 68.1 & \bnum{0.89} & 76.4 \\
    \midrule

    \multicolumn{5}{c}{\textbf{NAR / Continuous AR}} \\
    \midrule
    F5-TTS         & \xmark  & 2.00 & 67.0 & 1.53 & 76.0 \\
    VibeVoice      & \xmark  & 3.04 & 68.9 & 1.16 & 74.4 \\
    VoxCPM         & \xmark  & \bnum{1.85} & \bnum{72.9} & \bnum{0.93} & \bnum{77.2} \\
    \midrule

    \multicolumn{5}{c}{\textbf{Discrete AR}} \\
    \midrule
    Llasa         & \xmark    & 2.97 & 57.4 & 1.59 & 68.4 \\
    SparkTTS      & \xmark    & 1.98 & 58.4 & 1.20 & 67.2 \\
    OpenAudio-s1-mini& \xmark & 1.94 & 55.0 & 1.18 & 68.5 \\
    HiggsAudio-v2 & \xmark    & 2.44 & 67.7 & 1.50 & 74.0 \\
    FireRedTTS2   & \xmark    & 1.95 & 66.5 & \bnum{1.14} & 73.6 \\
    \rowcolor{oursgray}
    \textbf{CAT-TTS (Ours)} & \cmark &
    \bnum{1.89} & \bnum{73.1} &
    1.23 & \bnum{78.5} \\
    \bottomrule
  \end{tabular}
  % \vspace{-50pt }
\end{table}

\paragraph{Comparison with Open-Source TTS Systems.}

We evaluate the performance of our CAT-based fully autoregressive (AR) TTS system against a comprehensive suite of open-source models. These baselines encompass three major paradigms: (i) \textit{cascaded systems} (e.g., AR+NAR), (ii) \textit{purely non-autoregressive systems}, and (iii) \textit{prior purely autoregressive systems} based on discrete or continuous representations. Detailed descriptions and categorizations of these baseline systems are provided in Appendix \ref{appendix:baseline_tts}.

As shown in Table~\ref{table:tts_seed_compare}, CAT-TTS significantly outperforms previous discrete fully autoregressive models, particularly in speaker similarity (SIM). Moreover, our method achieves competitive performance compared to recent state-of-the-art systems such as IndexTTS2, MaskGCT, and VoxCPM, with all systems maintaining very low word error rates (WER), typically below 2\%.

Notably, CAT-TTS achieves the highest speaker similarity scores on Seed-TTS-Eval for both English and Chinese among the compared open-source models. This demonstrates that scaling CAT, as a unified discrete interface, effectively captures fine-grained acoustic characteristics required for high-quality, zero-shot speech generation.

\subsection{Speech Understanding}

In addition to speech generation, we further evaluate the speech understanding capability of CAT by applying it to downstream LLM-based ASR and comparing against representative open-source state-of-the-art speech understanding models; detailed results are provided in Appendix~\ref{appendix:speech_understanding_on_tac}.

\section{Analysis Of Scalling Behavior}
\label{sec:ablation}
\subsection{End-to-End Optimization Makes CAT a Scalable Audio Tokenizer}
\begin{figure}[htbp]
  \centering
  \includegraphics[width=0.9\linewidth]{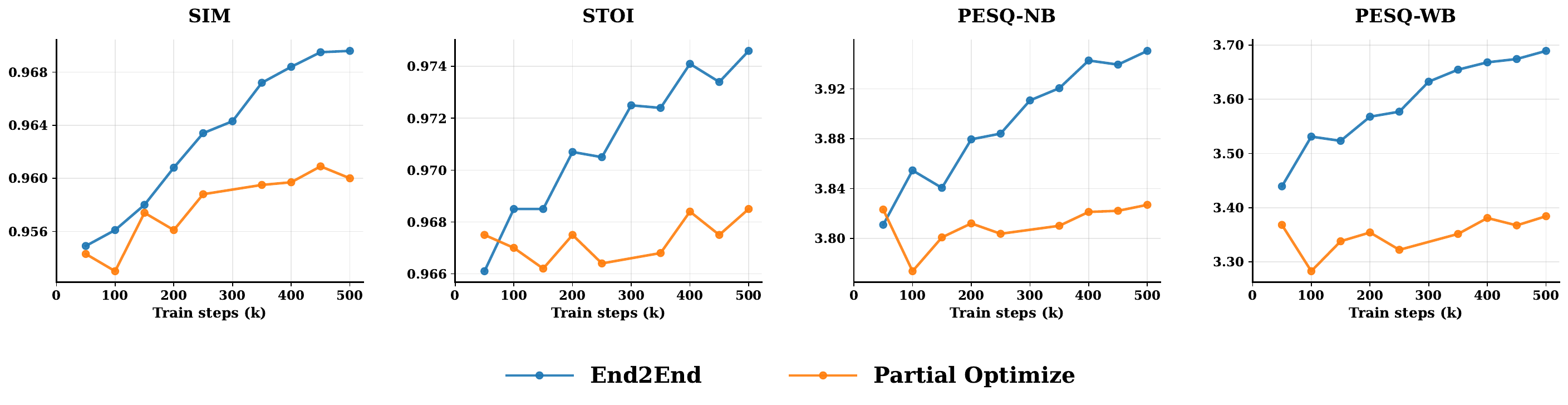}
\caption{Comparison between full end-to-end optimization and partial (stage-wise) optimization for CAT.}
\vspace{-6pt}
  \label{fig:ablation_end2end_vs_partial}
\end{figure}

A key goal of CAT is \textit{scalability}—the ability to continuously improve reconstruction quality with increased training budget. Although CAT consists of multiple adversarial components, its optimization strategy is critical for enabling such scalability. We compare \textbf{full end-to-end} optimization with the \textbf{partial} protocol used in prior works~\citep{wu2023audiodec, li2025baichuan, gong2025xy, zhang2025mimo}, where the encoder and quantizer are frozen while the decoder and discriminator are optimized.
As shown in Figure~\ref{fig:ablation_end2end_vs_partial}, end-to-end training yields sustained improvements across all metrics without early saturation. In contrast, partial optimization plateaus early, as freezing components restricts the model's ability to refine representations. These results demonstrate that \textbf{end-to-end optimization is crucial for scaling CAT effectively} with increased computation and capacity.

\label{sec:end2end}
\vspace{-8pt}

\subsection{Co-Scaling of Model Parameters and Quantization Capicity}
\label{sec:capacityscale}
\begin{figure}[h]
  \centering
  \includegraphics[width=1.0\linewidth]{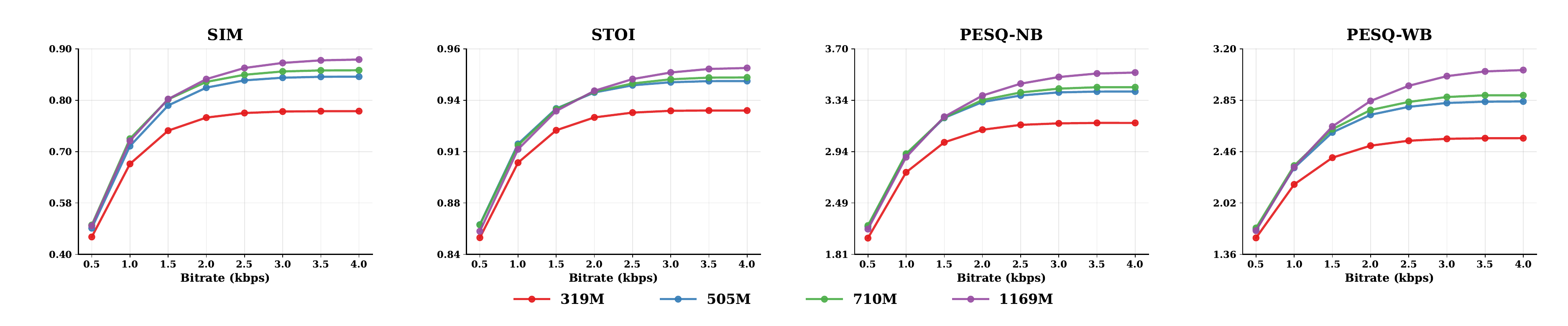}
\caption{Scaling behavior of CAT reconstruction performance with respect to bitrate and model parameters.}
  \label{fig:TAC_scale_dim}
\vspace{-6pt}
\end{figure}

We examine how CAT scales with model size. Following Section~\ref{main:method}, we jointly optimize all components while varying the hidden dimension (256, 384, 512, 768)—totaling 319M, 505M, 710M, and 1169M combined encoder--decoder parameters, respectively. Throughout, the quantizer is held constant at 32 layers and 12.5~Hz.

Figure~\ref{fig:TAC_scale_dim} shows that increasing the parameter count improves reconstruction quality across 0.5--4~kbps. While the 1169M model benefits most from high bitrates, smaller versions saturate early. Notably, at low bitrates, the 1169M model can underperform smaller models operating at higher bitrates, indicating that bitrate—rather than parameter count—becomes the primary bottleneck. These findings reveal that \textbf{parameter scaling and quantization depth are fundamentally co-dependent}. Neither can be scaled effectively in isolation, as system performance is governed by the narrowest bottleneck. Optimal scaling thus requires a synchronized expansion of both model parameters and quantization capacity within an end-to-end framework.

\begin{figure*}[h]
  \centering
  \includegraphics[width=1.0\linewidth]{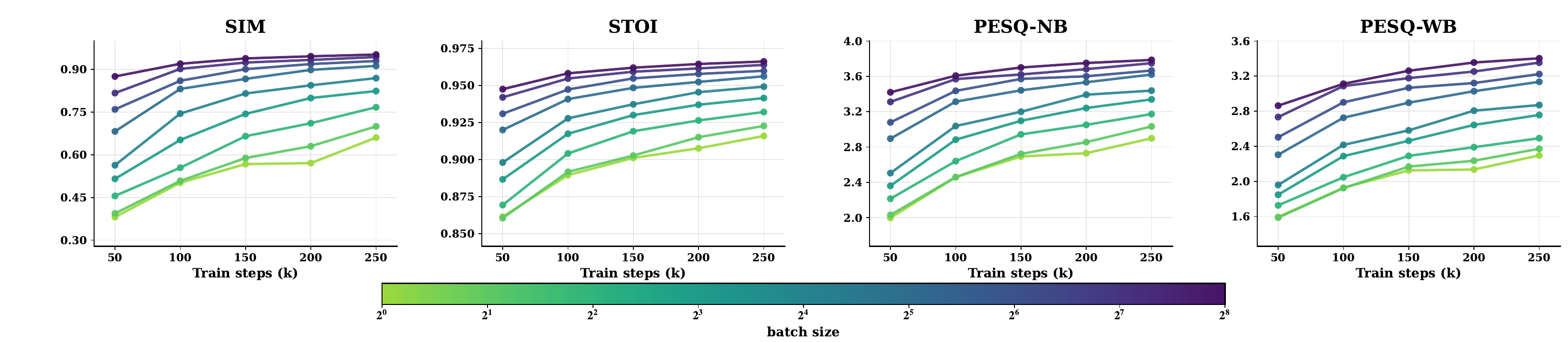}
  \caption{Scaling behavior of CAT reconstruction performance with respect to training batch size. The color gradient represents the batch size scale from $2^0$ to $2^8$. Larger batch sizes consistently yield superior reconstruction fidelity across all metrics.}
  \label{fig:TAC_scale_batch_size}
\end{figure*}

\subsection{Reconstruction Fidelity Benefits Consistently from Increased Training Scale}
\label{sec:batchsize_scale}

Beyond model parameters and bitrate, a key indicator of a tokenizer's scalability is its capacity to effectively translate increased training computation into higher fidelity. We investigate this by varying the global training batch size from a baseline factor of $2^0$ up to $2^8$, while keeping the total number of training steps and other hyperparameters constant.

As illustrated in Figure~\ref{fig:TAC_scale_batch_size}, CAT exhibits a clear and positive correlation between training scale and reconstruction quality across all evaluated speech metrics (SIM, STOI, and PESQ). At any given point in the training process, increasing the batch size yields strictly better performance. Notably, the performance curves for larger batch sizes maintain a strong upward trajectory even at 250k steps, achieving substantially higher quality within the same step budget compared to smaller scales.

These results suggest that CAT exhibits stable and predictable scaling behavior with respect to training batch size. Increasing data throughput leads to systematically higher-fidelity representations, highlighting the model's suitability for large-scale audio tokenizer training where computational resources can be traded directly for reconstruction quality.

\section{Related Works}
\label{related_works}
\subsection{Discrete Audio Tokenizers}

Discrete audio tokenizers aim to encode continuous audio waveforms into sequences of discrete tokens and reconstruct audio signals from these tokens. Most existing methods adopt an RVQGAN-style framework, which employs an encoder–quantizer–decoder architecture combined with adversarial training to achieve high-fidelity audio reconstruction~\citep{zeghidour2021soundstream,defossez2022high,kumar2023high,defossez2024moshi}.

SoundStream~\citep{zeghidour2021soundstream} introduces \textit{quantizer dropout}, enabling a single tokenizer to support variable bitrate reconstruction. Encodec~\citep{defossez2022high} further improves reconstruction quality by incorporating a multi-scale STFT (MS-STFT) discriminator to capture audio structures at different temporal resolutions. DAC~\citep{kumar2023high} simplifies the training process via factorized vector codes and employs complex STFT discriminators at multiple time scales to enhance phase modeling. Other acoustic codecs, including BigCodec~\citep{xin2024bigcodec}, Stable-Codec~\citep{parker2024scaling} and TS3-Codec~\citep{wu2024ts3}, focus on improving reconstruction quality under extremely low bitrates.

Beyond reconstruction fidelity, recent studies have explored injecting semantic information into audio tokenizers to better support downstream generative and understanding tasks. A common approach is knowledge distillation from pretrained teacher models~\citep{hsu2021hubert,chung2021w2v,chen2022wavlm}. SpeechTokenizer~\citep{zhang2023speechtokenizer}, Mimi~\citep{defossez2024moshi}, and Qwen3 TTS Tokenizer~\citep{hu2026qwen3} align the encoder and quantizer representations with self-supervised speech models through distillation objectives. In contrast, XCodec2.0~\citep{ye2025codec}, Higgs Audio Tokenizer~\citep{higgsaudio}, Dual Codec~\citep{li2025dualcodec}, and SAC~\citep{chen2025sac} directly initialize the tokenizer encoder using pretrained SSL or ASR models, thereby reducing the difficulty of semantic modeling.

A scale-driven approach introduces semantic information into audio tokenizers through large-scale audio–text supervision. Methods such as Baichuan Audio Tokenizer~\citep{li2025baichuan}, XY-Tokenizer~\citep{gong2025xy}, and MiMo Audio Tokenizer~\citep{zhang2025mimo} leverage audio-to-text tasks and massive paired datasets, enabling the tokenizer to implicitly learn rich semantic representations while maintaining high-fidelity reconstruction.

Despite these advances, it remains unclear what characteristics make an audio tokenizer truly suitable for native audio language models. We argue that such a tokenizer should minimize handcrafted priors and architectural constraints, and instead adopt a simple and scalable design. Our goal is to obtain an audio tokenizer that is well aligned with the modeling needs of audio language models by scaling up both computation and data and training the tokenizer in an end-to-end manner.

\subsection{Audio Generation}

Audio generation models have witnessed rapid progress in recent years~\citep{kreuk2022audiogen,borsos2023audiolm,liu2023audioldm,huang2023make}, largely driven by the combination of discrete audio representations~\citep{chung2021w2v,defossez2022high,zhu2025muq} and large-scale language modeling~\citep{kaplan2020scaling,achiam2023gpt}. A dominant paradigm is to perform generation in a compressed acoustic space, where audio is represented by sequences of discrete tokens produced by neural audio codecs~\citep{defossez2022high,defossez2024moshi}, and generation is formulated as a language modeling problem.

AudioLM~\citep{borsos2023audiolm} proposes a hierarchical generation strategy that decomposes audio generation into three stages: semantic modeling, coarse acoustic modeling, and fine acoustic modeling. By combining representations from self-supervised speech models~\citep{chung2021w2v} with neural codec tokens, AudioLM achieves high-quality audio generation with strong long-term consistency. VALL-E~\citep{wang2023neural} introduces a hybrid autoregressive (AR)~\citep{radford2018improving} and non-autoregressive (NAR)~\citep{devlin2019bert} architecture for speech synthesis, and demonstrates that scaling training data to tens of thousands of hours leads to the emergence of in-context learning capabilities for speech generation. Tortoise-TTS~\citep{betker2023better} further explores expressive text-to-speech by combining autoregressive sequence modeling with diffusion-based~\citep{ho2020denoising} refinement, enabling multi-voice and highly expressive synthesis.

Along this line, an important trend is the move toward end-to-end audio generation~\citep{agostinelli2023musiclm,liu2023audioldm,liao2024fish,ning2025diffrhythm,peng2025vibevoice}, where a single generative model directly produces audio tokens, rather than cascading multiple generative models (e.g., BERT-style or GPT-style language models, or diffusion-based generative models) in a multi-stage pipeline. This simplification substantially reduces system complexity and error propagation across stages, while also improving training stability and inference efficiency.

In the context of discrete token-based generation, MusicGen~\citep{agostinelli2023musiclm} systematically studies different multi-sequence modeling patterns and finds that the delay pattern enables a single autoregressive model to perform both text- and melody-conditioned music generation. More recent systems such as Moshi~\citep{defossez2024moshi} adopts a combination of temporal transformers and depth transformers to efficiently model long audio sequences, and further leverage streaming audio tokenizers to significantly reduce inference latency, enabling faster and more responsive audio generation.

Beyond discrete tokenization, there is also a growing body of work on audio generation based on continuous representations. These approaches augment auto-regressive large language models with local diffusion transformers (LocDiT)~\citep{liu2024autoregressive,jia2025ditar,peng2025vibevoice,zhou2025voxcpm}, enabling the auto-regressive model to directly generate continuous latent representations and capture fine-grained acoustic details without explicit discretization.

Overall, modern audio generation research is converging toward scalable, end-to-end architectures that tightly couple representation learning and generation. This trend highlights the increasing importance of well-designed audio tokenizers that are not only faithful in reconstruction quality, but also compatible with the architectural choices and scaling properties of audio language models~\citep{zhang2023speechtokenizer,defossez2024moshi,yuan2025yue,zhang2025mimo}.

\subsection{End-to-End Audio Language Models}

End-to-end audio language models~\citep{zhang2023speechgpt,nguyen2025spirit,defossez2024moshi,zeng2024glm,li2025baichuan,zhao2025moss,zhang2025mimo} aim to unify speech understanding, generation, and reasoning within a single large-scale model, moving beyond conventional three-stage pipelines that decompose speech processing into ASR, text-based language modeling, and TTS. By directly modeling audio representations using language modeling objectives, these systems aim to equip large language models with native audio understanding and generation capabilities.

Early efforts in this direction include SpeechGPT~\citep{zhang2023speechgpt}, which is among the first large-scale models to support end-to-end speech interaction. SpeechGPT leverages discrete speech representations derived from self-supervised speech encoders~\citep{hsu2021hubert} and scales training on large amounts of cross-modal data, enabling large language models to acquire intrinsic conversational abilities across speech and text modalities. Subsequent works such as Spirit-LM~\citep{nguyen2025spirit}, GLM4-Voice~\citep{zeng2024glm}, and MOSS-Speech~\citep{zhao2025moss} further improve speech–text alignment by scaling up speech–text interleaved data, demonstrating that tightly coupled multimodal pretraining is critical for robust end-to-end speech understanding and generation.

More recent systems push this paradigm to significantly larger scales. Models such as Kimi-Audio~\citep{ding2025kimi} and Qwen3-Omni~\citep{qwen3omni} expand training data to hundreds of thousands or even millions of hours of audio, leading to substantially improved robustness in complex and diverse audio scenarios. These results suggest that end-to-end audio language models benefit strongly from data scaling, similar to trends observed in text-only large language models~\citep{kaplan2020scaling,henighan2020scaling,achiam2023gpt}.

An emerging line of work explores end-to-end audio language modeling based on information-preserving or near-lossless audio representations~\citep{defossez2022high,defossez2024moshi}. Moshi~\citep{defossez2024moshi} employs a multi-stream speech-to-speech Transformer together with a streaming audio tokenizer to enable full-duplex spoken dialogue, achieving low-latency, highly responsive, and human-like interactions. MiMo-Audio~\citep{zhang2025mimo} further demonstrates that scaling training data to the order of 100 million hours allows end-to-end audio language models to exhibit emergent few-shot in-context learning capabilities in audio, highlighting the strong interaction between tokenizer design, data scale, and model capacity.

Overall, these studies highlight the central role of audio tokenizers in end-to-end audio language models. Similar to text tokenizers for LLMs, an audio tokenizer is expected to provide a native discrete interface that scales effectively with autoregressive modeling. Accordingly, our goal is to develop a unified, fully end-to-end trained audio tokenizer built from homogeneous causal Transformers, supporting predictable scaling with data and model capacity while minimizing handcrafted constraints.

% \section{Conclusion}
% We introduced \textbf{CAT}, 
% a fully end-to-end Transformer-based audio tokenizer designed as a unified discrete interface for autoregressive modeling. By jointly optimizing its components within a homogeneous, causal architecture, CAT achieves state-of-the-art reconstruction across speech, sound, and music while demonstrating predictable scaling behavior. Our results in zero-shot speech generation and understanding show that CAT effectively bridges the gap between raw audio and language modeling, providing a scalable and high-fidelity foundation for the next generation of native audio foundation models.

\section{Conclusion}
In this paper, we introduced \textbf{CAT} (\textbf{C}ausal \textbf{A}udio \textbf{T}okenizer with \textbf{T}ransformer), a fully end-to-end Transformer-based architecture that serves as a unified discrete interface for autoregressive audio language modeling. Leveraging the CAT architecture, we developed \textbf{MOSS-Audio-Tokenizer}, a 1.6-billion-parameter audio tokenizer pre-trained from scratch on 3 million hours of diverse audio data, effectively acquiring general audio representations across various domains.
Through the joint end-to-end optimization of all components—including the encoder, quantizer, decoder, discriminators, and a decoder-only LLM for semantic alignment—within a purely causal framework, MOSS-Audio-Tokenizer achieves state-of-the-art reconstruction performance among open-source audio tokenizers. Furthermore, its discrete representations demonstrate exceptional performance in both downstream speech generation and speech understanding. Our findings position the CAT architecture as a unified, scalable interface for the next generation of native audio foundation models.

% \newpage
\section*{Contributors}

\textbf{Contributors} \\
Yitian Gong$^{*}$, Kuangwei Chen, Zhaoye Fei, Xiaogui Yang, Ke Chen, Yang Wang, Kexin Huang, Mingshu Chen, Ruixiao Li, Qinyuan Cheng, Shimin Li

\vspace{0.8em}

\textbf{Advisors} \\
Xipeng Qiu$^{\dagger}$

\vspace{1.0em}

\noindent\textbf{Affiliations}: \\
MOSI Intelligence\\
Shanghai Innovation Institute\\
Fudan University

% 使用你要求的无编号脚注方法
{\let\thefootnote\relax\footnotetext{
    $^*$ \email{ytgong24@m.fudan.edu.cn} \quad 
    $^{\dagger}$ Corresponding author: \email{xpqiu@fudan.edu.cn}
}}

% \section*{Contributors}

% \textbf{Contributors} \\
% Yitian Gong, Kuangwei Chen, Zhaoye Fei, Xiaogui Yang, Ke Chen, Yang Wang, Kexin Huang, Mingshu Chen, Ruixiao Li, Qinyuan Cheng, Shimin Li

% \vspace{0.8em}

% \textbf{Advisors} \\
% Xipeng Qiu$^{\dagger}$

% \vspace{1.0em}

% \noindent\textbf{Affiliations}: \\
% MOSI Intelligence\\
% Shanghai Innovation Institute\\
% Fudan University

% {\small
% % $^{*}$ Equal contribution. \\
% $^{\dagger}$ Corresponding author.
% }

% ===== Bibliography =====
\clearpage
\bibliographystyle{unsrtnat}
\bibliography{main}

% ===== Appendix (content unchanged) =====
\clearpage
\beginappendix

\startcontents[app]
\begingroup
  \renewcommand{\contentsname}{Appendix Contents}
  \section*{\contentsname}
  \printcontents[app]{}{1}{}
\endgroup
\newpage

\section{More Details Of MOSS-Audio-Tokenizer}
\label{appendix:implementation_details}
\subsection{Architecture}

The encoder and decoder of MOSS-Audio-Tokenizer each consist of 68 causal Transformer blocks with a 10\,s sliding-window attention, enabling efficient streaming inference.
To progressively reduce the sequence length, the encoder inserts patchify operations~\citep{dosovitskiy2020image} at the input stage and after layers 12, 24, and 36, with patch sizes of 240/2/2/2, respectively. Since patchify operations modify the feature dimensionality, a linear projection is applied after each patchify stage to map the hidden states to the corresponding Transformer block dimension.
This design maps raw 24\,kHz waveforms to a low frame rate of 12.5\,Hz.

The encoder is composed of four stages with hidden dimensions of 768, 768, 768, and 1280, respectively.
These stages contain 12, 12, 12, and 32 Transformer blocks.
For each stage, the feed-forward network (FFN) dimension is set to four times the corresponding hidden dimension.
Multi-head self-attention uses 12, 12, 12, and 20 attention heads for the four stages, respectively.
All Transformer blocks employ rotary positional embeddings (RoPE)~\citep{su2024roformer}.

The decoder mirrors the encoder architecture in a fully causal manner.
Both the encoder and decoder contain approximately 0.8B parameters and are trained from scratch.

% The encoder and decoder of CAT each consist of 68 causal Transformer blocks with a 10\,s sliding-window attention, enabling streaming inference. To progressively reduce sequence length, the encoder inserts patchify operations at the input stage and after layers 12, 24, and 36, with patch sizes 240/2/2/2, mapping 24\,kHz waveforms to a low frame rate of 12.5\,Hz. 这些 transfomrer block 的维度分别是 768. 768. 768. 1280, 分别有 12, 12, 12, 32 层， FFN dim 是各自 dim 的 4 倍，attention heads 分别是 12， 12， 12，20。都使用 ROPE 位置编码 The decoder mirrors the encoder causally. Both the encoder and decoder contain approximately 0.8B parameters and are trained from scratch.

Discrete tokenization is performed using a 32-layer residual vector quantizer (RVQ). Each layer uses a codebook of size 1024 with factorized vector quantization (latent dimension 8)~\citep{kumar2023high} and  L2-normalized codes. Quantizer dropout with probability 1.0 is applied during training to enable variable-bitrate tokenization.

To encourage semantically structured discrete representations, we attach a 0.5B decoder-only causal language model~\citep{qwen2.5} for audio-to-text supervision, which autoregressively predicts text conditioned on the quantizer outputs. The audio-to-text tasks include ASR, multi-speaker ASR, and audio captioning.

For adversarial training, we employ a multi-period discriminator~\citep{defossez2022high} and a complex STFT discriminator~\citep{kumar2023high}. All components—encoder, quantizer, decoder, semantic head, and discriminators—are optimized jointly in an end-to-end manner.

\subsection{Dataset and Optimization.}
We train MOSS-Audio-Tokenizer on approximately 3M hours of speech, sound, and music data, covering both clean and in-the-wild recordings, and mixing audio-only and paired (audio, text) samples. For samples with available transcriptions or captions, we apply an auxiliary audio-to-text training objective, while audio-only samples are used without text supervision. We optimize both the generator and discriminators using AdamW~\citep{loshchilov2017decoupled} optimizer and conduct training in bfloat16 (bf16) precision. The generator is trained with a learning rate of $1\times10^{-4}$ and a weight decay of 0.01, while no weight decay is applied to the discriminators. The loss weights are set to $\lambda_{\mathrm{sem}}{=}20$, $\lambda_{\mathrm{rec}}{=}15$, $\lambda_{\mathrm{cmt}}{=}0.25$, $\lambda_{\mathrm{code}}{=}1.0$, $\lambda_{\mathrm{adv}}{=}1.0$, and $\lambda_{\mathrm{feat}}{=}2.0$.

\subsection{Training schedule.}
Due to computational constraints, we adopt a two-stage training schedule to improve training efficiency:
 non-adversarial pretraining without discriminator-related losses for 520k steps (batch size 1536, approximately 5 hours of audio per batch), followed by adversarial finetuning for 500k steps (batch size 768). All modules are optimized end-to-end without pretrained encoders or semantic teachers~\citep{hsu2021hubert,radford2023robust,zhang2023speechtokenizer,defossez2024moshi,ye2025codec}.

\section{More Details on Evaluation of Audio Tokenizers}
\label{appendix:evaluation_audio_tokenizer}

\subsection{Reconstruction Evaluation Protocol}

We evaluate the reconstruction quality of MOSS-Audio-Tokenizer and open-source audio tokenizers across three domains: \textit{speech}, \textit{sound}, and \textit{music}.

\paragraph{Objective evaluation.}
For speech reconstruction, we conduct evaluations on LibriSpeech test-clean (English)~\citep{panayotov2015librispeech} and AISHELL-2 (Chinese)~\citep{du2018aishell}.
We report speaker similarity (SIM), computed as the cosine similarity between speaker embeddings extracted from the original and reconstructed audio using a pretrained speaker verification model\footnote{\url{https://github.com/microsoft/UniSpeech/tree/main/downstreams/speaker_verification}}.
In addition, we report short-time objective intelligibility (STOI)~\citep{taal2010short} and perceptual evaluation of speech quality (PESQ)~\citep{rix2001perceptual}.

For sound and music reconstruction, following prior work~\citep{kumar2023high}, we evaluate on the AudioSet evaluation subset~\citep{gemmeke2017audio} and MUSDB~\citep{rafii2017musdb18}.
We report mel-spectrogram distance and short-time Fourier transform (STFT) distance as objective metrics.

\paragraph{Subjective evaluation.}
In addition to objective metrics, we conduct a crowd-sourced listening test based on the MUSHRA protocol~\citep{series2014method}.
In this test, each listener rates the perceptual quality of reconstructed audio samples on a 1--100 scale.

For tokenizers that support variable bitrate decoding, we report results at multiple bitrates to characterize reconstruction quality across different bitrate regimes.

\subsection{Results Of Subjective Evaluation}
% \begin{figure}[h]
%   \centering
%   \includegraphics[width=1\linewidth]{Images/mushra_speechen_v2.pdf}
%     \caption{
%     Comparison of the MUSHRA subjective test results between open-source audio tokenizers and CAT.
%     }
%   \label{fig:Mushra_results}
% \end{figure}

\begin{wrapfigure}{r}{0.5\textwidth}
\vspace{-12pt}
  \centering
  \includegraphics[width=\linewidth]{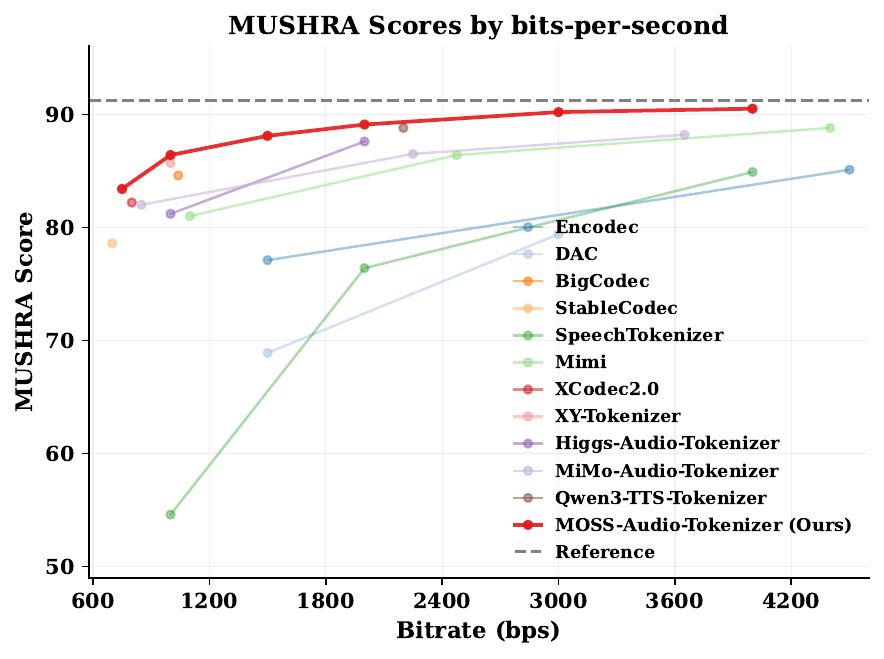}
  \caption{MUSHRA subjective evaluation results.}
  \label{fig:Mushra_results}
  \vspace{-12pt}
\end{wrapfigure}

We conduct subjective evaluations on speech data to compare MOSS-Audio-Tokenizer with open-source audio tokenizers.
For tokenizers that support variable bitrates, we report subjective scores at multiple bitrates.
The results are shown in Figure~\ref{fig:Mushra_results}.

Overall, MOSS-Audio-Tokenizer achieves strong and consistent performance across a wide range of bitrates, indicating high perceptual quality in reconstructed speech.
For Encodec, DAC, and SpeechTokenizer, the subjective scores are competitive at higher bitrates but degrade noticeably at lower bitrates.
In contrast, audio tokenizers designed for a specific target bitrate (e.g., BigCodec, XCodec~2.0, XY-Tokenizer, and the Qwen3 TTS Tokenizer) perform well at their respective training bitrates, where their perceptual quality is competitive with MOSS-Audio-Tokenizer at comparable bitrates.

% 我们在 speech 数据上对 CAT 和开源 audio tokenizer 做了主观评测, 对于支持可变 bps 的 audio tokenizer, 我们测了多个比特率下的主观指标. results are shown in Figure~\ref{fig:Mushra_results}. 我们发现 CAT 在多个比特率下都表现的很好, 这说明 CAT 重建音频的感知质量较好, For Encodec, DAC, SpeechTokenizer, 他们在高 bps 下结果较好,但是低 bps 下结果下降明显. For audio tokenizer designed for a 特定比特率(例如 BigCodec, XCodec2.0, XY-Tokenizer, Qwen3 TTS Tokenizer), 他们在他们训练的比特率下表现得较好, competitve to CAT 在相近 bps. 

Overall, these results demonstrate that MOSS-Audio-Tokenizer provides a scalable and robust tokenizer for general audio, enabling high-fidelity compression and reconstruction of speech, sound, and music across a wide range of bitrates.

\subsection{Baseline Audio Tokenizers}
% \label{appendix:baseline_audio_tokenizers}

In this section, we provide additional implementation and configuration details for the baseline audio tokenizers reported in Table~\ref{table:landscape_of_different_audio_tokenizers}. Unless otherwise specified, for models based on vector quantization, the target bitrate is controlled during evaluation by truncating residual vector quantization (RVQ) codes to the first several layers.

\paragraph{Encodec.}
We evaluate the official causal EnCodec model operating at 24\,kHz for monophonic audio\footnote{\url{https://huggingface.co/facebook/encodec_24khz}}~\citep{defossez2022high} and it contains approximately 14\,M parameters.

\paragraph{DAC (Descript Audio Codec).}
DAC~\citep{kumar2023high} is a neural audio codec designed for high-fidelity waveform reconstruction using carefully engineered discriminators and improved vector quantization strategies. We use the official 24\,kHz monophonic model for evaluation. The released checkpoint contains approximately 74\,M parameters.

\paragraph{SpeechTokenizer.}
We adopt the official \texttt{speechtokenizer\_hubert\_avg} model\footnote{\url{https://huggingface.co/fnlp/SpeechTokenizer/tree/main/speechtokenizer_hubert_avg}}, which is trained on monophonic speech at 16\,kHz~\citep{zhang2023speechtokenizer}. SpeechTokenizer distills HuBERT representations using the first layer of residual vector quantization, enabling effective disentanglement of speech information, and further supports a unified speech language model (USLM). The model contains approximately 103.67\,M parameters.

\paragraph{Mimi.}
We evaluate the official Mimi codec\footnote{\url{https://huggingface.co/kyutai/mimi}}~\citep{defossez2024moshi}. Mimi operates on monophonic audio at 24\,kHz and produces discrete audio tokens at a frame rate of 12.5\,Hz, while supporting streaming encoding and decoding.

\paragraph{BigCodec.}
We use the authors' released checkpoint with the default 16\,kHz monophonic configuration. BigCodec~\citep{xin2024bigcodec} employs a single vector quantization (VQ) codebook with a size of 8{,}192 and produces discrete tokens at an 80\,Hz frame rate. The model contains approximately 159\,M parameters.

\paragraph{Stable Codec.}
For Stable Codec, we use the released \texttt{stable-codec-speech-16k-base} checkpoint\footnote{\url{https://huggingface.co/stabilityai/stable-codec-speech-16k-base}}, which operates on monophonic speech at 16\,kHz. Stable Codec~\citep{parker2024scaling} adopts a residual finite scalar quantization (RFSQ) bottleneck. Following the official recommendation, we apply the \texttt{1x46656\_400bps} and \texttt{2x15625\_700bps} FSQ bottleneck preset during evaluation. The base checkpoint contains approximately 953\,M parameters.

\paragraph{XCodec2.0.}
XCodec2.0 is a semantically enhanced speech codec that incorporates a pre-trained speech encoder~\citep{chung2021w2v}. We use the authors' released checkpoint\footnote{\url{https://huggingface.co/HKUSTAudio/xcodec2}} and follow the official inference pipeline. XCodec2.0 encodes 16\,kHz monophonic audio into discrete tokens at a 50\,Hz frame rate using a single-layer vector quantizer. The released checkpoint contains approximately 822\,M parameters.

\paragraph{XY-Tokenizer.}
XY-Tokenizer~\citep{gong2025xy} is designed to mitigate the semantic--acoustic conflict at ultra-low bitrates by jointly modeling semantic and acoustic information using two encoders. We evaluate the officially released checkpoint\footnote{\url{https://huggingface.co/fdugyt/XY_Tokenizer}}. XY-Tokenizer encodes 16\,kHz monophonic audio into discrete tokens at a 12.5\,Hz frame rate using an 8-layer RVQ (codebook size 1{,}024). Quantizer dropout is disabled in the released model. The tokenizer contains approximately 519\,M parameters.

\paragraph{Higgs Audio Tokenizer.}
We evaluate the released \texttt{Higgs-audio-v2-tokenizer} checkpoint\footnote{\url{https://huggingface.co/bosonai/higgs-audio-v2-tokenizer}}~\citep{higgsaudio}, which operates on monophonic audio at 24\,kHz. The checkpoint used in our experiments contains approximately 201\,M parameters.

\paragraph{MiMo Audio Tokenizer.}
MiMo-Audio-Tokenizer~\citep{zhang2025mimo} is designed to support both waveform reconstruction and downstream language modeling. The tokenizer jointly optimizes semantic and reconstruction objectives on a large-scale corpus, reportedly exceeding 11 million hours of audio. In our evaluation, we use the official released checkpoint\footnote{\url{https://huggingface.co/XiaomiMiMo/MiMo-Audio-Tokenizer}}. The model contains approximately 1.2\,B parameters.

\paragraph{Qwen3 TTS Tokenizer.}
Qwen3-TTS-Tokenizer~\citep{hu2026qwen3} is the discrete speech tokenizer used in Qwen3-TTS for speech generation and streaming text-to-speech. We evaluate the released tokenizer checkpoint\footnote{\url{https://huggingface.co/Qwen/Qwen3-TTS-Tokenizer-12Hz}} on monophonic audio at 24\,kHz. The tokenizer encodes waveforms into discrete tokens at a frame rate of 12.5\,Hz and contains approximately 170\,M parameters.

\begin{figure}[t]
  \centering
  \includegraphics[width=0.5\linewidth]{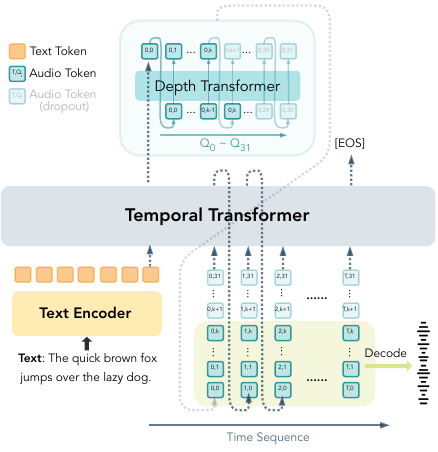}
\caption{
Architecture of bitrate controllable audio modeling.
During training, Progressive Sequence Dropout randomly truncates the number of active RVQ layers.
During inference, when decoding with a fixed depth $k$, only the first $k$ RVQ tokens are provided as input at each time step, and the Depth Transformer autoregressively predicts only these $k$ tokens, while finer RVQ layers are omitted.
}
  \label{fig:tts_arch}
\end{figure}

\section{More Details Of Bitrate Controllable Speech Generation}
\label{appendix:speech_generation}

% \begin{wrapfigure}{r}{0.50\textwidth}
% \vspace{-14pt}
%   \centering
%   \includegraphics[width=\linewidth]{Images/TTS_arch.pdf}
% \caption{
% Architecture of bitrate controllable audio modeling.
% During training, Progressive Sequence Dropout randomly truncates the number of active RVQ layers.
% During inference, when decoding with a fixed depth $k$, only the first $k$ RVQ tokens are provided as input at each time step, and the Depth Transformer autoregressively predicts only these $k$ tokens, while finer RVQ layers are omitted.
% }
%   \label{fig:tts_arch}
% \vspace{-14pt}
% \end{wrapfigure}

\subsection{Architecture}
The Temporal Transformer is initialized from Qwen3-1.7B~\citep{yang2025qwen3}.
The Depth Transformer is randomly initialized and consists of 4 Transformer blocks with a hidden size of 1536 and an FFN dimension of 8960.

\subsection{Training Details}

We adopt a global batch size of 1.35M tokens, including both text tokens and speech tokens, where speech tokens are counted at a frame rate of 12.5\,Hz.
During training, text tokens and speech tokens are concatenated along the temporal dimension to form the input sequence for the TTS model.
FlashAttention-2~\citep{dao2023flashattention} is used to accelerate training.
All models are trained using the AdamW optimizer with a peak learning rate of $2\times10^{-4}$.
For ablation studies, models are trained for 50k steps, while the final models are trained for 200k steps.

\subsection{Inference Details}

At inference time, to maintain consistency with the training procedure, we synthesize speech in a continuation-based manner.
Given a prompt audio with its transcription and a target text to be synthesized, we concatenate the prompt transcription tokens $\mathbf{x}_{\mathrm{prompt}}$, the target text tokens $\mathbf{x}_{\mathrm{syn}}$, and the prompt audio tokens $\mathbf{s}_{\mathrm{prompt}}$ into a single input sequence.
The TTS model then autoregressively predicts the speech tokens corresponding to the target text.
Finally, the predicted speech tokens are decoded into waveforms using the CAT decoder.

% \subsection{Training Details}
% We adopt a global batch size of 1.35M tokens, including both text tokens and speech tokens, where speech tokens are counted at a frame rate of 12.5\,Hz. 训练时 text token 和 speech token 在时间维度上 concat 起来后作为 TTS 系统的训练数据。 训练时使用 Flash Attention2 加速
% All models are trained using the AdamW optimizer with a peak learning rate of $2\times10^{-4}$.
% For ablation studies, models are trained for 50k steps, while the final models are trained for 200k steps.

% \subsection{Inference Details}
% 推理时，为了和训练时的行为保持一致，我们采用续写的方式来合成音频，给定 prompt 音频和 prompt 转录文本，以及要合成的音频对应的文本，我们将 prompt 转录文本 token x_{prompt}, 要合成的音频对应的文本 token x_{syn}, 以及 prompt 音频 token s_{prompt} concat 后的 token 序列作为 TTS 模型的输入，让他自回归的预测待生成的语音 token，最后将语音 token 送到 CAT decoder 中得到语音。

\section{More Details on Baseline Text-to-Speech Systems}
\label{appendix:baseline_tts}

We compare our CAT-TTS system with a wide range of open-source text-to-speech (TTS) models.
These models can be broadly categorized into three groups.

The first group consists of cascaded TTS systems that employ multiple generative models, such as AR+NAR or NAR+NAR architectures.
Representative examples include MaskGCT~\citep{wang2024maskgct}, FireRedTTS~\citep{xie2025fireredtts}, CosyVoice2~\citep{du2024cosyvoice}, Qwen2.5-Omni~\citep{qwen2.5omni}, CosyVoice3~\citep{du2025cosyvoice}, IndexTTS2~\citep{zhou2025indextts2}, FlexiCodec-TTS~\citep{li2025flexicodec}, and GLM-TTS~\citep{cui2025glm}.

The second group includes purely non-autoregressive TTS systems, such as F5-TTS~\citep{chen2025f5}.

The third group comprises prior fully autoregressive TTS models based on either discrete or continuous representations, including Llasa~\citep{ye2025llasa}, SparkTTS~\citep{wang2025spark}, OpenAudio-s1~\citep{openaudios1}, HiggsAudio-v2~\citep{higgsaudio}, FireRedTTS2~\citep{xie2025fireredtts}, DiTAR~\citep{jia2025ditar}, and VoxCPM~\citep{zhou2025voxcpm}.

CAT-TTS adopts a purely autoregressive architecture based on discrete tokens to perform zero-shot TTS in an end-to-end manner, directly generating speech from text without relying on predefined intermediate representations, such as semantic tokens~\citep{hsu2021hubert,du2024cosyvoice}.
Moreover, CAT-TTS supports variable-bitrate speech generation through Progressive Sequence Dropout.

\section{Speech Understanding On CAT}
\label{appendix:speech_understanding_on_tac}
\definecolor{oursgray}{gray}{0.92}

\begin{table}[h]
  \centering
  \fontsize{8.5pt}{11pt}\selectfont
  \renewcommand{\arraystretch}{1.05}
  \caption{ASR performance comparison.} % Caption 也做了简化
  \label{table:asr_simplified}
  \resizebox{.45\textwidth}{!}{
  \begin{tabular}{cccc}
    \toprule
     & \textbf{Model Size} & \textbf{EN-WER} $\downarrow$ & \textbf{ZH-CER} $\downarrow$ \\
    \midrule
    Whisper-Large-v3  &  1.5B  & 2.90 & 5.80 \\
    Voxtral Small-24B  &  24B & 1.53 & 13.80 \\
    FireredASR-AED   &   1.1B  & 1.93 & 3.00 \\
    Qwen2-Audio-Base  &   7B & 1.74 & 3.08 \\
    Baichuan-Audio-Base  & 7B & 3.02 & 3.87 \\
    Step-Audio-Chat  &   130B  & 3.11 & 3.60 \\
    Qwen2.5-Omni    &    7B  & 2.37 & 2.56 \\
    Kimi-Audio      &    7B & 1.28 & 2.56 \\
    \rowcolor{oursgray}
    \textbf{CAT-ASR (Ours)} & 1.7B & \textbf{2.96} & \textbf{3.44} \\
    \bottomrule
  \end{tabular}
  }
\end{table}
We explore the capability of CAT for speech understanding tasks by developing \textbf{CAT-ASR}.
Specifically, we investigate whether CAT tokens can be directly used as inputs to a large language model (LLM) for automatic speech recognition (ASR), in order to evaluate the alignment between CAT and text as well as the information preservation of the discrete speech representation.

We adopt Qwen3-1.7B~\citep{yang2025qwen3} as the backbone LLM.
To enable speech understanding, we initialize a set of 32 speech tokens in the vocabulary and directly feed the discretized CAT speech tokens into the LLM.
For each speech frame, the tokens along the RVQ dimensions are summed and treated as a single input embedding to the LLM.
The model is then trained in a fully autoregressive manner to predict the corresponding text sequence given the speech token inputs.

The model is trained on an internal dataset consisting of approximately 2 million hours of paired $(\text{audio}, \text{text})$ data.
We use a global batch size of 1M tokens and train the model for 200k steps with a warmup of 4k steps.
The Adam optimizer is adopted with a peak learning rate of $5\times10^{-5}$.
All experiments are conducted without any additional alignment or auxiliary supervision beyond the standard ASR objective.

We evaluate the trained CAT-based ASR model on both English and Chinese benchmarks.
For English, we report word error rate (WER) on the LibriSpeech \texttt{test-clean} set~\citep{panayotov2015librispeech}.
For Chinese, we report character error rate (CER) on the AIShell-2 iOS subset~\citep{du2018aishell}.
We compare our model with a range of previous open-source ASR systems and speech-language models~\citep{radford2023robust,liu2025voxtral,xu2025fireredasr,chu2024qwen2,li2025baichuan,huang2025step,qwen2.5omni,ding2025kimi}, as summarized in Table~\ref{table:asr_simplified}.

As shown in Table~\ref{table:asr_simplified}, CAT-ASR achieves competitive performance across both English and Chinese benchmarks.
These results suggest that CAT tokens retain sufficient linguistic content and exhibit good alignment with text, enabling effective speech understanding when directly consumed by an LLM.
We believe CAT-ASR can be further improved by scaling up paired training data and model capacity.

\end{document}